\documentclass[twocolumn]{autart}    
\usepackage{cite}
\usepackage{bm}
\usepackage[export]{adjustbox}
\usepackage{caption}
\usepackage{subcaption}
\usepackage{graphicx}          
\usepackage{mathtools}                               
\usepackage{multicol}
\usepackage{multirow}
\usepackage{wrapfig}
\let\theoremstyle\relax
\usepackage{amsthm}
\usepackage{amsmath}
\usepackage{amsfonts}
\usepackage{amssymb}
\usepackage{makecell}
\usepackage{marginnote}

\newtheorem{theorem}{Theorem}[section]
\theoremstyle{definition}

\newtheorem{assumption}[theorem]{\textbf{Assumption}}
\newtheorem{lemma}[theorem]{Lemma}

\usepackage{algorithm, algpseudocode, xspace}
\usepackage[english]{babel}
\usepackage[utf8]{inputenc}
\usepackage{graphicx}
\usepackage{accents}
\usepackage{enumitem}
\usepackage{tikz}
\usetikzlibrary{positioning}
\algnewcommand{\parState}[1]{\State%
	\parbox[t]{\dimexpr\linewidth-\algmargin}{\strut\hangindent= \algorithmicindent \hangafter=1 #1\strut}}

\usepackage{array}
\usepackage{pdfpages}
\usepackage{tikz}
\usetikzlibrary{positioning}
\usepackage{mathtools}
\allowdisplaybreaks
\newcolumntype{P}[1]{>{\centering\arraybackslash}p{#1}}
\begin{document}

\begin{frontmatter}

\title{Federated reinforcement learning for robot motion planning with zero-shot generalization\thanksref{footnoteinfo}} 

\thanks[footnoteinfo]{This work was partially supported by NSF grants ECCS-1710859, CNS-1830390, ECCS-1846706 and the Penn State College of Engineering Multidisciplinary Research Seed Grant Program.}

\author{Zhenyuan Yuan}\ead{zqy5086@psu.edu},    
\author{Siyuan Xu}\ead{spx5032@psu.edu},               
\author{Minghui Zhu}\ead{muz16@psu.edu}  

\address{School of Electrical Engineering and Computer Science, \\The Pennsylvania State University, University Park, PA., 16802, USA}  


\begin{abstract}                          
This paper considers the problem of learning a control policy for robot motion planning with zero-shot generalization, i.e., no data collection and policy adaptation is needed when the learned policy is deployed in new environments. 
We develop a federated reinforcement learning framework that enables collaborative learning of multiple learners and a central server, i.e., the Cloud, without sharing their  raw data. In each iteration, each learner uploads its local control policy and the corresponding estimated normalized arrival time to the Cloud, which then computes the global optimum among the learners and broadcasts the optimal policy to the learners. Each learner then selects between its local control policy and that from the Cloud for next iteration. The proposed framework leverages on the derived zero-shot generalization guarantees on arrival time and safety. Theoretical guarantees on almost-sure convergence, almost consensus, Pareto improvement and optimality gap are also provided. Monte Carlo simulation is conducted to evaluate the proposed framework.
\end{abstract}

\end{frontmatter}

\section{Introduction}
Robotic motion planning is a fundamental problem that allows robots to execute a sequence of actions and achieve certain tasks, such as reaching goal regions and grasping objects. Classic motion planning methods usually assume perfect knowledge of the dynamics of the robots and the environments they operate in. Examples of methods includes cell decomposition, roadmap,  sampling-based approaches, and feedback motion planning. Interested readers are referred to \cite{lavalle2006planning} for more details. However, robots' operations in the real world are usually accompanied by uncertainties, such as the external disturbances in the natural environments they operate in and the  modeling errors  of the dynamics.  
To deal with the uncertainties, a number of methods utilize techniques in robust control (e.g., \cite{danielson2020robust,majumdar2017funnel,lakshmanan2020safe}), where bounded uncertainties are considered, and stochastic control (e.g., \cite{omainska2021gaussian,ono2015chance,castillo2020real}), where the uncertainties are modeled in terms of known probability distributions.
Recently, reinforcement learning-based approaches
have been developed to relax the need of prior explicit uncertainty models (and even the dynamic models) by directly learning the best mapping from  sensory data to  control inputs from repetitive trials. For example, paper \cite{virani2018imitation} uses kernel methods  to learn the control policy for a spider-like robot with 18 degrees of freedom using GPS data. 
Deep neural networks are used in  \cite{kantaros2022perception,fu2019neural} to synthesize control policies  using camera/LiDAR data.

	Classic reinforcement learning problems consider learning an optimal control policy over a single environment \cite{sutton2018reinforcement}. The policy can either be learned online through agent's repetitive interaction and data collection in the environment \cite{sutton2018reinforcement} or learned offline using a fixed dataset of past interaction with the environment \cite{levine2020offline}.
Although the methods can deal with complex environments, the agents struggle to generalize their experiences to new environments \cite{cobbe2019quantifying,kirk2023survey}.
This paper focuses on the generalization of reinforcement learning, that is, obtaining a control policy which performs well in new environments unseen during training. 
	Depending on whether or not the approaches require  data collection and policy adaptation in a new environment, existing works on this problem can be categorized into few-shot generalization and zero-shot generalization.

Meta reinforcement learning (MRL) is a widely-used approach for few-shot generalization. More specifically,
	MRL aims to address the fundamental problem of  quickly learning an optimal control policy in a new environment after collecting a small amount of data online and performing a few updates for policy adaptations \cite{finn2017model,pong2022offline,ji2020convergence,rakelly2019efficient, sun2021provably,xu2022meta}. The problem is usually formulated as an  optimization problem, where the objective function is the expected performance of the  control policy  adapted from a meta control policy after a few updates in a new environment.  However, as pointed out by \cite{kirk2023survey}, for safety purpose a control policy still needs to be reasonably good at deployment time (i.e. zero-shot) even if the policy continues learning during deployment. Furthermore, when it is applied to robots with unknown dynamics, MRL faces a particular challenge. Since they usually operate in real time, robots only have limited time to collect data in new environments and perform policy adaptation. When the dynamics of the robots are uncertain, data collection requires that the robots execute the meta control policies in physical environments and obtain the induced trajectories. The physical execution can be time-consuming and not suitable or even impractical for real-time applications.

Zero-shot generalization considers the performance of a single control policy in new environments  without additional data collection and policy adaptation  \cite{kirk2023survey}. It is typically formulated as expected cost minimization of a control policy over a distribution of environments.
As the distribution of the environments is generally complicated or even unknown, it is challenging, if not impossible, to solve the expected cost minimization problem in closed form. Therefore, the methods, which target zero-shot generalization, instead solve an empirical mean minimization problem (possibly with regularization) given a finite amount of training environments.   Related methods can be categorized into two classes. The first one is modifying an expected cost function and solving the modified problem through empirical cost minimization \cite{garcia2015comprehensive,schaul2015universal,moldovan2012risk,gosavi2009reinforcement,heger1994consideration,nilim2005robust}. For example, risk-sensitive criterion can be introduced to balance between a return and a risk, where the risk can be the variance of the return \cite{moldovan2012risk,gosavi2009reinforcement}.  Worst-case criterion is used to mitigate the effects of the variability induced by a given policy due to the stochastic nature of the unseen environments or the dynamic  systems \cite{heger1994consideration,nilim2005robust}.
The other class is incorporating regularizers into empirical mean minimization to improve the generalizability of the solution. 
A necessarily incomplete list of references includes \cite{majumdar2018pac,lenz2015deep,levine2016end,mahler2018dex}. While most regularization methods are heuristic,  paper \cite{majumdar2018pac} uses the sum of the empirical cost and the generalization error from PAC-Bayes theory as an upper bound of the expected cost and synthesizes a control policy which can minimize the upper bound. Nevertheless, empirical mean minimization (with regularization)  is an approximation to the expected cost minimization problem, and the optimality loss is not quantified. 
In this paper, we aim to directly solve the expected cost minimization problem and analyze the properties of the solution.



The papers aforementioned  focus on centralized reinforcement learning, where all the training data are possessed by a single learning agent. On the other hand, the advent of ubiquitous sensing and mobile storage renders some scenarios, in which training data are distributed across multiple entities, e.g., the driving data in different autonomous cars. It is well-known that  control policies trained with more data have better performance \cite{vapnik2013nature}. However,  directly using the raw data for collective learning can risk compromising the privacy of the data owners, e.g., exposing the living and working locations of the drivers.
To tackle this challenge, distributed reinforcement learning is usually leveraged, where multiple learning agents perform training collaboratively by exchanging their locally learned models. There are mainly two approaches: decentralized reinforcement learning and federated reinforcement learning. 
In decentralized reinforcement learning, learning agents directly communicate with each other over P2P networks \cite{zhang2018fully}.  In federated reinforcement learning, learning agents cannot directly talk to each other and instead are orchestrated by a Cloud, i.e., the learning agents download shared control policies from the Cloud, implement local updates based on local data and report the local control policies to the Cloud for the updates of the shared models \cite{fan2021fault,khodadadian2022federated}. 
With the support of a Cloud, federated learning has access to more resources in, e.g., computation, memory and power, and hence enables a much larger scale of learning processes. The analysis of the above works is limited to the convergence of the proposed learning algorithms. The generalization of the learned control policies remains an open question.

{\bf Contribution statement: } 
In this paper, we propose a novel framework, FedGen, to tackle the challenge of robot motion planning with zero-shot generalization in the presence of distributed data across multiple learning entities. 
A network of learners aim to collaboratively learn a single control policy which can safely drive a robot to goal regions in different environments without data collection and policy adaptation during policy execution.  The problem is formulated as federated optimization with an unknown objective function, which is the expected cost of navigation over a distribution of environments. Specifically, each learner updates its local control policy and sends its observation of the objective function to a central Cloud for global minimization among the control policies of the learners. The global minimizer is then sent back to the learners for updates of the local control policies.  We characterize the upper bounds for the expected arrival time and safe arrival rate for each control policy. The upper bounds are used to find the control policy with the best zero-shot generalization performance among the learners.  Theoretical guarantees on almost-sure convergence, almost consensus, Pareto improvement and optimality gap are also provided. In addition, the algorithm can be executed over P2P networks after a minor change.

In summary, our contributions are:
(C1) The development of the FedGen algorithm for robot motion planning with zero-shot generalization subject to multiple learning entities.
(C2) The theoretic guarantees on the zero-shot generalization of local control policy to new environments in terms of arrival time and safety, the almost-sure convergence and the optimality gap of the local estimates, the consensus of the local values and Pareto improvement of the local values.
Monte Carlo simulations are conducted for evaluations. 

{\bf Distinction statement.} Compared to the preliminary version \cite{yuan2023federated}, this paper provides a new Theorem \ref{thm: global convergence}, which characterizes the optimality  gap. Table \ref{tab: performance vs learner number} presents new simulation results comparing the performances of the algorithm with respect to different numbers of learners.  Furthermore, Section \ref{sec: proofs} includes all the proofs of the theoretical results, which are omitted in \cite{yuan2023federated} due to space limitation. In addition, in Section \ref{sec: discussion}, we provide discussions on hyperparameter tuning, the trade-off in the selection of a hyperparameter as well as the effects on the optimality of the control policies in terms of the number of learners and the sample sizes in the learners.

{\em Notations.} We use superscript $(\cdot)^{[i]}$ to distinguish the local values of robot $i$ and $\|\cdot\|$ to denote  2-norm. For notional simplicity, for any local value $a^{[i]}$, we denote $a^{\max}\triangleq \max_{i\in\mathcal{V}}a^{[i]}$ and $a^{\min}\triangleq\min_{i\in\mathcal{V}}a^{[i]}$. Define closed ball $\mathcal{B}(\theta,\epsilon)\triangleq\{\theta'\in\mathbb{R}^{n_\theta}\mid \|\theta-\theta'\|\leqslant\epsilon\}$,  and $\beta(\mathcal{A})$ the measure of set $\mathcal{A}$.

\section{Problem Formulation}
In this section, we introduce the dynamics of the robot, the problem of motion planning, the setting of federated  reinforcement learning, and the objective of this paper. 

\subsection{Environment-specific motion planning} 
In this paper, we consider environment-dependent dynamics.  Let $\mathcal{X}\subseteq\mathbb{R}^{n_x}$ be the state space of the robot and $\mathcal{U}\subseteq\mathbb{R}^{n_u}$ be the control input space. 
An environment $E$ is fully specified by the inherent external disturbance  $d_E:\mathcal{X}\times\mathcal{U}\to\mathcal{X}$, the obstacle region $\mathcal{X}_{O,E}\subseteq \mathcal{X}$ and the goal region $\mathcal{X}_{G,E}\subset\mathcal{X}\setminus\mathcal{X}_{O,E}$; i.e., $E\triangleq(d_E, \mathcal{X}_{O,E},\mathcal{X}_{G,E})$.  For each environment $E$, denote free region $\mathcal{X}_{F,E}\triangleq\mathcal{X}\setminus \mathcal{X}_{O,E}$. Denote $\mathcal{G}_{\mathcal{E}}$ the space of goal regions induced by the space of environments $\mathcal{E}$. 
	
	In each environment $E$, the dynamic system of the robot is given by the following difference equation:
\begin{equation}\label{eq: dynamic model}
	x_{t+1}=f(x_t,u_t)+d_E(x_t,u_t),~
	o_t=h(x_t,\mathcal{X}_{O,E}),
\end{equation}
where $x_t\in\mathcal{X}$ is the state of the robot, $u_t\in\mathcal{U}$ is its control input, $o_t\in\mathcal{O}$ is the sensor output of the system observing the obstacle region $\mathcal{X}_{O,E}$ at state $x_t$ and $h$ is the observation function. Once environment $E$ is revealed, $\mathcal{X}_{G,E}$ is known, $\mathcal{X}_{O,E}$ can only be observed through $h$ and may not be fully known, but $d_E$ is unknown.

The objective of the environment-specific motion planning problem is to synthesize a control policy, which can drive system \eqref{eq: dynamic model} to the goal region with obstacle collision avoidance. 
The arrival time under control policy $\pi:\mathcal{O}\times\mathcal{G}_{\mathcal{E}}\to\mathcal{U}$  for system \eqref{eq: dynamic model} starting from initial state $x_{int}$ is given by 
\begin{align*}
	t_E&(x_{int};\pi)\triangleq\inf\{t>0\mid x_t\in\mathcal{X}_{G,E}, x_0=x_{int}, \\
	&x_{\tau+1}=f(x_\tau,u_\tau)+d_E(x_\tau,u_\tau),~o_\tau=h(x_\tau,\mathcal{X}_{O,E}),\\
	&u_\tau=\pi(o_\tau;\mathcal{X}_{G,E}),x_\tau\in\mathcal{X}_{F,E}, \forall 0\leqslant\tau\leqslant t\}.    
\end{align*}
If the robot never reaches the goal, or hits the obstacles before arrival, then $t_E(x_{int};\pi)=\infty$. We say safe arrival is achieved from initial state $x_{int}$ under control policy $\pi$ if $t_E(x_{int};\pi)<\infty$.
Note that $t_E(x_{int};\pi)$ is potentially infinite,  and it can cause numerical issues. Therefore, we 
normalize the arrival time function through Kruzkov transform such that the normalized cost function is given by $J_E(x_{int};\pi)\triangleq1-e^{-t_E(x_{int};\pi)}$. Note that when $t_E(x_{int;\pi})=\infty$, we have $J_E(x_{int};\pi)=1$.

\subsection{Robot motion planning with zero-shot generalization}
In the problem of robot motion planning with zero-shot generalization, the goal is to synthesize a single control policy that performs well in different environments without data collection and policy adaptation during policy execution. In statistical learning theory \cite{vapnik2013nature}, this can be formulated as minimizing the expectation of the normalized arrival time over different environments.
In particular,  we assume the environments follow an unknown distribution.
\begin{assumption}\label{assmp: stochastic environment}
	{\em(Stochastic environment).}
	There is an unknown distribution $\mathcal{P}_E$ over $\mathcal{E}$ from which environments are drawn from.$\hfill\blacksquare$
\end{assumption}

For example, the obstacle regions of the environments can be composed of a number of circular obstacles, where  the numbers, locations, and the radii of the obstacles follow an unknown distribution, and the disturbances can follow an unknown Gaussian process.

Further, we assume that the initial state is a random variable which is conditional on the environment.
\begin{assumption}\label{assmp: stochastic initialization}
	\emph{(Stochastic initialization).} There is an unknown conditional distribution $\mathcal{P}_{int|E}$ from which $x_{int}$ is drawn  conditional on  environment $E\in\mathcal{E}$.
	$\hfill\blacksquare$
\end{assumption}

Formally, the objective of the problem of robot motion planning with zero-shot generalization is to synthesize a control policy $\pi_*\in\Gamma\triangleq \{u(\cdot): \mathcal{O}\times\mathcal{G}_{\mathcal{E}}\to\mathcal{U},\textrm{measurable}\}$, such that the expected normalized cost over all possible, including unseen, environments is minimized:
\begin{align}\label{obj:arrival time}
	\pi_*=\arg\min_{\pi\in\Gamma} \mathbb{E}[J_{E}(x_{int};\pi)], 
\end{align}
where the expectation is taken over the environment $E\sim\mathcal{P}_E$ and initialization $x_{int}\sim\mathcal{P}_{int\mid E}$. Note that by taking the expectation, we are considering all possible environments following the distribution. Therefore, we measure the zero-shot generalization of a control policy using its expected cost of solving the motion planning problems in a distribution of environments.

Since $\Gamma$ is a function space, problem \eqref{obj:arrival time} is a functional optimization problem and hard to solve in general. In order to make the problem tractable, we  approximate the space $\Gamma$ using, e.g., deep neural networks and basis functions.
Consider a class of control policies $\pi_{\theta}\in\Gamma$ parameterized by $\theta\in\mathbb{R}^{n_{\theta}}$, e.g., the weights of a deep neural network. 
Denote $\eta(\theta)\triangleq \mathbb{E}[J_{E}(x_{int};\pi_{\theta})]$. Then for the learners, problem \eqref{obj:arrival time} becomes:
\begin{align}\label{obj:prameterize}
	\theta_*=\arg\min_{\theta\in\mathbb{R}^{n_{\theta}}} \eta(\theta). 
\end{align}
Problem \eqref{obj:prameterize} is a standard expected cost minimization problem. 
However, since the distribution of the environments is unknown, \eqref{obj:prameterize} cannot be solved directly. A typical practice is to approximate it by empirical cost minimization (with regularization), e.g.,
\cite{moldovan2012risk,gosavi2009reinforcement,abe2010optimizing,heger1994consideration,nilim2005robust,majumdar2018pac,lenz2015deep,levine2016end,mahler2018dex}, where a control policy is synthesized by minimizing the empirical cost (with regularization) over a finite number of training environments. Nevertheless, to the best of our knowledge, there is no theoretic guarantee on the optimality of the solutions to the original problem \eqref{obj:prameterize}.
 In this paper, we aim to directly solve \eqref{obj:prameterize} and  analyze  the properties of the solutions.

\subsection{Federated reinforcement learning}\label{sec: fed learning}

Through federated learning, a group of learners aim to solve \eqref{obj:prameterize} collaboratively and achieve better results than solving on their own.
 Each learner $i\in\mathcal{V}$ observes function $\eta$ by sampling a set of  environments  $E^{[i]}_l\stackrel{i.i.d.}{\sim}\mathcal{P}_E$, $l=1,\cdots,n_{\mathcal{E}}^{[i]}$, and a set of initial states $x^{[i]}_{int|E^{[i]}_{l},l'}\sim\mathcal{P}_{int|E^{[i]}_{l}}$, $l'=1,\cdots,n_{int|\mathcal{E}}^{[i]}$, for each $E^{[i]}_l$. We consider general on-policy reinforcement learning methods.
Given a triple of $(\theta^{[i]}, E^{[i]}_l, x^{[i]}_{int|E^{[i]}_l,l'})$, learner $i$ measures the value $J_{E^{[i]}_l}(x^{[i]}_{int|E^{[i]}_l,l'};\pi_{\theta^{[i]}})$  through policy evaluation, i.e.,  running the robot under control policy $\pi_{\theta^{[i]}}$  from initial state $x^{[i]}_{int|E^{[i]}_l,l'}$ in  environment $E^{[i]}_l$, measuring the arrival time and taking the Kruzkov transform.
Then learner $i$ finds (or approximate using, e.g., natural evolution strategies \cite{wierstra2014natural}) the policy gradient $\nabla_{\theta^{[i]}} J_{E^{[i]}_l}(x^{[i]}_{int|E^{[i]}_l,l'};\pi_{\theta^{[i]}})$.  
 The learners communicate to a Cloud but do not communicate with each other.

The objective of the multi-learner network and the Cloud is to collaboratively solve problem \eqref{obj:prameterize}.
The problem is  challenged by the fact that the objective function $\eta$ is  non-convex and can only be estimated by sampling over the environments and the initial states in general. As stated in Assumption \ref{assmp: stochastic environment}, the environments at training and testing follow an unknown distribution. The estimation error is the difference between the true value of $\eta$ and the empirical average of the normalized cost, and the distribution of the estimation error  is unknown and non-Gaussian in general. 
Notice that when expected cost minimization is approximated by empirical cost minimization (possibly with regularization) as in \cite{moldovan2012risk,gosavi2009reinforcement,abe2010optimizing,di2012policy,heger1994consideration,nilim2005robust,majumdar2018pac,lenz2015deep,levine2016end,mahler2018dex}, the surrogate objective function is the sum of the empirical cost and the regularizer, which has closed-form and is free of estimation error. 

\begin{algorithm}[t]
	\caption{FedGen}\label{alg:collaborative optimization}
	\begin{algorithmic}[1]
		\State \textbf{Input: } Local sample sizes: $n_{\mathcal{E}}^{[i]}, n_{int|\mathcal{E}}^{[i]}$;  Kruzkov transform constant: $\alpha$; Initial step size: $r^{[i]}$; Initial estimate: $\theta^{[i]}_0$; Threshold for gradient: $q^{[i]}$; Local bias: $b^{[i]}_\gamma$; Step exponent: $\rho\in(2/3,1)$.
		
		\State \textbf{Init: } 
		$\zeta^{[i]}_0\leftarrow 1$, 
		$\textsf{Stop}_0^{[i]}\leftarrow \textsf{False}$.
		\label{ln: init}
		

		\For{$k=1,2,\cdots, K$}
		
		\Statex \{Learner-based update\}
		\For{$i\in \mathcal{V}$}
		\If{ $\textsf{Stop}_{k-1}^{[i]}==\textsf{False}$}
		\State Collects $(y^{[i]}_{k-1},z^{[i]}_{k-1})$
		
		\EndIf
		
		
		\State Sends $(\theta^{[i]}_{k-1},y^{[i]}_{k-1})$ to the Cloud
		\If{$\|z^{[i]}_{k-1}\|\geqslant q^{[i]}$ and $\textsf{Stop}_{k-1}^{[i]}==\textsf{False}$}\label{ln: if z<q}
		
		\State 
		$\hat{\theta}_k^{[i]}\leftarrow\theta^{[i]}_{k-1}-\frac{r^{[i]}}{k^\rho} z^{[i]}_{k-1}$ \label{ln: update hat theta}

		
		\Else
		\State $\hat{\theta}_k^{[i]}\leftarrow\theta^{[i]}_{k-1}$ \label{ln: Else}
		
		\State $(y^{[i]}_{k},z^{[i]}_{k})\leftarrow ( y^{[i]}_{k-1}, z^{[i]}_{k-1})$\label{ln: y z = y z -}
		
		\State $\textsf{Stop}_k^{[i]}\leftarrow\textsf{True}$\label{ln: Converge}
		
		\EndIf
		
		\EndFor
		\Statex \{Cloud update\}

		\State $(j,l)\leftarrow \arg\min_{i\in\mathcal{V}, l'=0,\cdots, k-1}y^{[i]}_{l'}+b^{[i]}_\gamma$\label{ln: server global min}
		\State Sends $(\theta^{[j]}_l, y^{[j]}_{l}, b^{[j]}_\gamma)$ to all $i\in\mathcal{V}$
		\Statex \{Learner-based fusion\}
		\For{$i\in \mathcal{V}$}
		\If{$j\neq i$ and $y^{[j]}_{l}+b^{[j]}_\gamma<\min\{y^{[i]}_{k-1}-b^{[i]}_\gamma, \zeta^{[i]}_{k-1}\}$ and $\textsf{Stop}_{k-1}^{[i]}==\textsf{True}$}\label{ln: If eta < and xi >}
		\State  $\theta_k^{[i]}\leftarrow \theta^{[j]}_l$\label{ln: proj SV1}
		\State $\zeta^{[i]}_k\leftarrow y^{[j]}_{l}$\label{ln: zeta update}
		\State $\textsf{Stop}_k^{[i]}\leftarrow\textsf{False}$\label{ln: converge false}
		
		\Else 
		\State  $\theta_k^{[i]}\leftarrow\hat{\theta}^{[i]}_k$\label{ln: proj SV2}
		\State $\zeta^{[i]}_k\leftarrow \zeta^{[i]}_{k-1}$\label{ln: zeta = zeta -}
		\EndIf

		\EndFor
		
		\EndFor
		
	\end{algorithmic}
\end{algorithm}

\section{Algorithm Statement}
In this section, 
we propose a federated optimization framework, FedGen in Algorithm \ref{alg:collaborative optimization}, and analyze the generalized performances and the properties of the local estimates of the solution to problem \eqref{obj:prameterize} the algorithm renders.  Overall, the proposed solution enables learning with distributed data without data sharing.
The generalizability of a control policy is  characterized by an upper bound of $\eta$, the expected adjusted arrival time, using the empirical mean of the adjusted arrival time in Theorem \ref{thm: monotonic optimality}. We leverage the architecture of federated optimization, where the learners only exchange the parameters of their control policies and  minimize the above upper bound  to optimize the generalizability of its control policy. More detailed description of the proposed framework can be found in the subsection below.

\subsection{The FedGen algorithm}\label{sec: fed opt}
Denote $\theta^{[i]}_k$ the empirical estimate of the solution to problem \eqref{obj:prameterize} by learner $i$ at iteration $k$. Denote $y_k^{[i]}$, the empirical estimate of $\eta(\theta^{[i]}_k)$, and  $z_k^{[i]}$, the empirical estimate  of  $\nabla\eta(\theta^{[i]}_k)$ as follows. 
\begin{align*}
	y_k^{[i]} &\triangleq \frac{1}{n_{\mathcal{E}}^{[i]}n_{int|\mathcal{E}}^{[i]}}\sum_{l=1}^{n_{\mathcal{E}}^{[i]}}\sum_{l'=1}^{n_{int|\mathcal{E}}^{[i]}}J_{E^{[i]}_l}(x^{[i]}_{int|E^{[i]}_l,l'};\pi_{\theta_k^{[i]}}),\\
	z_k^{[i]}&\triangleq \frac{1}{n_{\mathcal{E}}^{[i]}n_{int|\mathcal{E}}^{[i]}}\sum_{l=1}^{n_{\mathcal{E}}^{[i]}}\sum_{l'=1}^{n_{int|\mathcal{E}}^{[i]}}\allowbreak \nabla J_{E^{[i]}_l}(x^{[i]}_{int|E^{[i]}_l,l'};\pi_{\theta_k^{[i]}}).
\end{align*}

\begin{figure}
	\centering
	\includegraphics[width=0.45\textwidth]{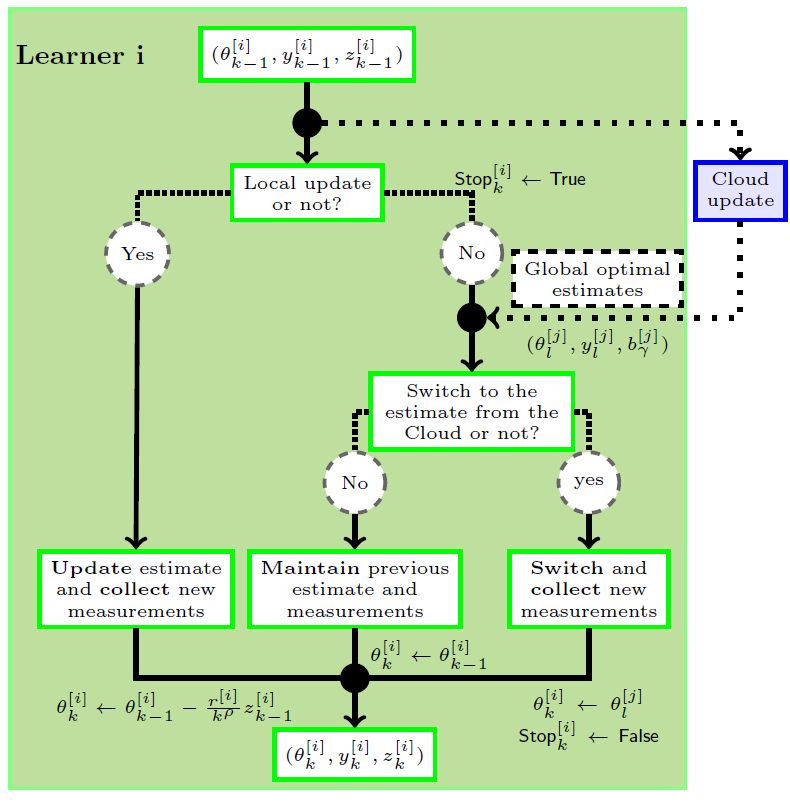}
	\caption{Implementation FedGen for learner $i$ in iteration $k$}
	\label{fig: flowchart}
\end{figure}

\begin{figure}
	\centering
	\includegraphics[width=0.45\textwidth]{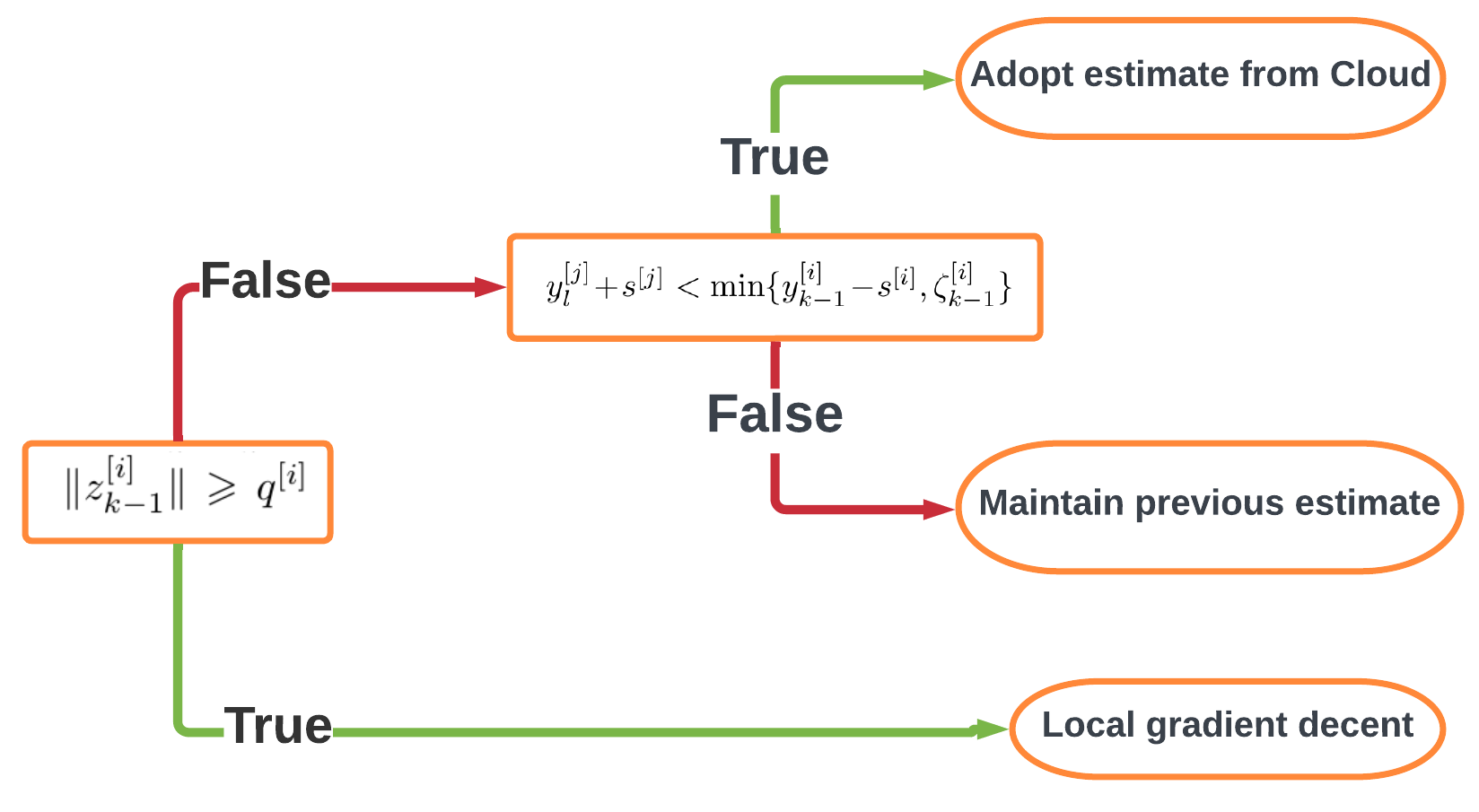}
	\caption{Parameter update logic at each iteration}
	\label{fig: logic}
\end{figure}
 

The FedGen algorithm is composed of three components: (i) Learner-based update, where each learner updates its estimate $\theta^{[i]}_k$ using local data only. (ii) Cloud update, where the Cloud identifies the estimate with the best generalized performance among the learners. (iii) Learner-based fusion, where the learner decides whether it should keep its local estimate or switch to the one returned by the Cloud. The algorithm utilizes the power of the Cloud to identify the control policy that can potentially achieve better performance in expectation and allow the learners to escape from their local minima. Figure \ref{fig: flowchart} is a detailed flowchart representation of Algorithm \ref{alg:collaborative optimization}, demonstrating the decision making process within learner $i$. Figure \ref{fig: logic} presents the logic of the update of the parameter estimates in one iteration. More detailed description of the each module in each iteration $k$ can be found below.

\subsubsection{Learner-based update}
First,  each learner $i$ performs local learning using its local data. Specifically, each learner $i$ collects the measurement $(y^{[i]}_{k-1},z^{[i]}_{k-1})$ of the estimate $\theta^{[i]}_{k-1}$ in the previous iteration if it is not stopped.  The measurements are sent to the Cloud for global minimization. 
 If  $\|z^{[i]}_{k-1}\|$ is greater than a local threshold $q^{[i]}$, which indicates that learner $i$'s estimate is far from convergence and has potential for improvement, the learner
makes one gradient descent step and updates its local estimate to $\hat{\theta}^{[i]}_k$. The threshold $q^{[i]}$ indicates whether a local minimum of $\eta$ is achieved.
  If $\|z^{[i]}_{k-1}\|$ is not greater than  $q^{[i]}$, the learner stops its local gradient descent and maintains the previous measurement. The learner resumes data collection for potential local gradient descent when it adopts the policy parameter from the Cloud later in Learner-based fusion for further optimization.

\subsubsection{Cloud update}
Note that the learners' estimates have different update trajectories due to the differences in initialization and data. Since objective $\eta$ is nonconvex in general, different learners' estimates can stuck at different local minima. Therefore, the Cloud aims to identify which learner is around a better local minimum such that the other learners can later switch to this local minimum when their estimates converges in Learner-based update. Specifically, upon the receipt of local estimates of $\eta$, $(y^{[i]}_{k-1}, \theta^{[i]}_{k-1})$, from each $i\in\mathcal{V}$,
the Cloud aims to find the policy parameter with the best generalized performance among the learners. Denote local bias $b^{[i]}_\gamma\triangleq\sqrt{\frac{\log (2/\gamma)}{2n_{\mathcal{E}}^{[i]}n_{int|\mathcal{E}}^{[i]}}}$, $\gamma\in(0,1)$. The following theorem characterizes the zero-shot generalization error between $y^{[i]}_k$ and $\eta(\theta^{[i]}_k)$ and the zero-shot generalized safety in terms of local bias, where the proof can be found in Section \ref{sec: proofs}.
\begin{theorem}\label{thm: monotonic optimality}
	Suppose Assumptions \ref{assmp: stochastic environment} and \ref{assmp: stochastic initialization}
	hold.  
	The following properties are true for all $i\in\mathcal{V}$:
	\begin{enumerate}
		\setlength{\itemindent}{-2em}
		\item[] \emph{(T1, Generalization error).}   For each $k\geqslant 0$, it holds that $\eta(\theta_k^{[i]})\leqslant y_k^{[i]}+ b^{[i]}_{\gamma}$ with probability at least $1-\gamma$.
		\item[] \emph{(T2, Generalized safety).} For each $k\geqslant 0$,  the policy $\pi_{\theta^{[i]}_k}$ is able to achieve safe  arrival   with probability at least $1-\gamma- (1-\gamma)(y^{[i]}_k+b^{[i]}_{\gamma})$ for $E\sim \mathcal{P}_E$ and $x_{int}\sim \mathcal{P}_{int|E}$.
		$\hfill\blacksquare$
	\end{enumerate}

	
	
\end{theorem}
In order to obtain the best zero-shot generalized performance, based on Theorem \ref{thm: monotonic optimality}, the Cloud
 returns the global minimizer of $y^{[i]}_{l'}+b^{[i]}_\gamma$ over  all the local estimates $\theta^{[i]}_{l'}$, $i\in\mathcal{V}, l'=0,\cdots, k-1$,  and sends the global minimizer and minimum to the learners.
Different from the regularizers used in the literature of empirical cost minimization, the local bias $b^{[i]}_\gamma$ is a constant value and does not depend on the estimate $\theta^{[i]}_k$.
This procedure can be implemented recursively by comparing the learner-wise global minimum in the previous iteration with the values obtained in the current iteration. If one wants to implement Algorithm \ref{alg:collaborative optimization} over P2P networks without the Cloud, this step can be executed using the minimum consensus algorithm \cite{olfati2002distributed}.

\subsubsection{Learner-based fusion}
For each learner, it may not be always the case that the global minimizer of the Cloud outperforms the local estimate. The learner's estimate only switches to the estimate returned from the Cloud if its estimate converges in Learner-based update and the estimate from the Cloud is significantly better than the local estimate. Specifically,
Learner $i$ only chooses  the global minimizer $\theta^{[j]}_{l}$ sent by the Cloud when two conditions are satisfied: (i)  estimate $\theta^{[j]}_{l}$ achieves a smaller estimate of $\eta$, i.e., $y^{[j]}_{l}+b^{[j]}_\gamma$ is less than the minimum between $y^{[i]}_{k-1}-b^{[i]}_\gamma$,  and $\zeta_{k-1}^{[i]}$, the previous global minimum  adopted by learner $i$; and (ii) local gradient descent is stopped, i.e.,   $z^{[i]}_{k-1}$ is small.
When the global minimizer is chosen, learner $i$ is then not stopped and  resumes Learner-based update in the next iteration.
 Notice that if it never chooses the global minimizer from the Cloud after it is stopped, learner $i$ maintains the estimate and measurement for the remaining iterations.

\subsection{Performance guarantees}\label{sec: guarantees}

In this section, we  investigate the limiting behavior of the algorithm. Similar to most analysis of stochastic gradient descent (please see \cite{fehrman2020convergence,ghadimi2013stochastic} and the references therein), we assume $\eta$ is Lipschitz continuous and $L_{\nabla\eta}$-smooth.
\begin{assumption}\label{assmp: lipschitz}
	\emph{(Lipschitz continuity).}
	There exists positive constant $L_\eta$  such that $|\eta(\theta)-\eta(\theta')|\leqslant L_\eta \|\theta-\theta'\|$ for all $\theta,\theta'\in\mathbb{R}^{n_{\theta}}$. $\hfill\blacksquare$
\end{assumption}
\begin{assumption}\label{assmp: smooth}
	\emph{($L_{\nabla\eta}$-smooth).}
	There exists positive constant $L_{\nabla\eta}$  such that $\|\nabla\eta(\theta)-\nabla\eta(\theta')\|\leqslant L_{\nabla\eta} \|\theta-\theta'\|$ for all $\theta,\theta'\in\mathbb{R}^{n_{\theta}}$. $\hfill\blacksquare$
\end{assumption}

Furthermore, we assume that  the variance of the errors of gradient estimation is bounded. This is a standard assumption in the analysis of stochastic optimization \cite{fehrman2020convergence}\cite{ghadimi2013stochastic}.
\begin{assumption}\label{assmp: ||z||}
	{\em (Bounded variance).}
	It holds that $\mathbb{E}[\|z^{[i]}_k-\nabla\eta(\theta^{[i]}_k)\|^2]\leqslant (\sigma^{[i]})^2$ for some $\sigma^{[i]}>0$. $\hfill\blacksquare$
\end{assumption}

Notice that the updates of the variables $\theta^{[i]}_k$, $y^{[i]}_k$ and $z^{[i]}_k$, $k\geqslant1$, depends on the sampling of the environments  and the initial states in all the learners, which are the only randomness in this paper.
	Therefore, in the sequel, {\em all} the expectations of these local variables  are taken over the sampling $E^{[j]}_l\sim \mathcal{P}_E$, $l=1,\cdots, n^{[j]}_\mathcal{E}$, and
 $x^{[j]}_{int\mid E^{[j]}_l, l'}\sim \mathcal{P}_{int\mid E^{[j]}_l}$, $l'=1,\cdots,n^{[j]}_{int\mid\mathcal{E}}$ for all $j\in\mathcal{V}$.
The lemma below shows that $z^{[i]}_k$ is an unbiased estimate of $\nabla\eta(\theta^{[i]}_k)$. 
\begin{lemma}\label{lemma: unbiased estimator of z}
{\em(Unbiased estimator).}
	Suppose Assumptions \ref{assmp: stochastic environment}, \ref{assmp: stochastic initialization} and \ref{assmp: lipschitz}  hold. Then it holds that $\mathbb{E}[z_k^{[i]}]-\nabla\eta(\theta_k^{[i]})=0$ for all $k\geqslant 1$. $\hfill\blacksquare$
\end{lemma}
Since $z^{[i]}_k$ is an unbiased estimate of $\nabla\eta(\theta^{[i]}_k)$, by the law of large numbers (Proposition 6.3 in \cite{bhattacharya2016course}), $(\sigma^{[i]})^2$ diminishes as $n^{[i]}_{\mathcal{E}}$ and $n^{[i]}_{int|\mathcal{E}}$ increase.

The following theorem summarizes the properties of almost-sure convergence, almost consensus and Pareto improvement of the algorithm. 
\begin{theorem}\label{thm: pareto improvement} 
	Suppose Assumptions \ref{assmp: stochastic environment}, \ref{assmp: stochastic initialization}, \ref{assmp: lipschitz}  \ref{assmp: smooth}  and  \ref{assmp: ||z||} hold. 
For all $i\in\mathcal{V}$,	if $r^{[i]}\leqslant \frac{1}{2L_{\nabla\eta}}$  and
	$q^{[i]}\geqslant 4\sigma^{[i]}$, then the followings hold:
	\begin{enumerate}
		\setlength{\itemindent}{-2em}
	\item[] \emph{(T3, Almost-sure convergence).}   There exists $\theta^{[i]}_\infty\in\mathbb{R}^{n_\theta}$ such that $\theta^{[i]}_k\to{\theta^{[i]}_\infty}$ almost surely.
	\item[] \emph{(T4, Almost consensus).} It holds that $\mathbb{E}[\max_{j\in\mathcal{V}} \allowbreak \eta(\theta^{[j]}_\infty)-\min_{j\in\mathcal{V}}\eta(\theta^{[j]}_\infty)]\leqslant 2b^{\max}_\gamma$.
		\end{enumerate}
Denote $k^{[i]}_{fs}\triangleq\min\{k\geqslant0\mid \|z^{[i]}_{k}\|<q^{[i]}\}$ the first time learner $i$ is stopped. Then we further have 
	\begin{enumerate}
		\setlength{\itemindent}{-2em}
	\item[] \emph{(T5, Pareto improvement).} If $\theta^{[i]}_\infty\neq\theta^{[i]}_{k^{[i]}_{fs}}$, then $\mathbb{E}[\eta(\theta^{[i]}_\infty)-\eta(\theta^{[i]}_{k^{[i]}_{fs}}) ]\leqslant-2b^{\min}_\gamma$. $\hfill\blacksquare$
	\end{enumerate}

\end{theorem}
Note that  $\theta^{[i]}_\infty\neq\theta^{[i]}_{k^{[i]}_{fs}}$ implies that learner $i$ adopts the estimates from the Cloud at least once.
Theorem \ref{thm: pareto improvement} (T5) implies that communication with the Cloud can potentially improve the optimality of the learners' estimates.

Denote the set of global minimizers  that are regular in the sense of Hurwitz as $$
\Theta_*\triangleq\{\theta\in\mathbb{R}^{n_\theta}\mid \theta=\arg\min_{\theta'\in\mathbb{R}^{n_\theta}}\eta(\theta'),\nabla^2\eta(\theta)\succ 0\}.
$$ 
Lemma 1 in \cite{mertikopoulos2020almost} indicates that for each $\theta_*\in\Theta_*$, there exists a convex compact neighborhood $\mathcal{K}(\theta_*)$ and constant $\alpha>0$ such that
\begin{align}\label{ineq: lemma 1 cited}
	\alpha\|\theta-\theta_*\|^2\leqslant \langle \nabla\eta(\theta),\theta-\theta_*\rangle, ~\forall\theta\in\mathcal{K}(\theta_*).
\end{align}
 Define $\epsilon_0(\theta_*)\triangleq \max\{\epsilon>0\mid \mathcal{B}(\theta_*,4\epsilon+2\sqrt{\epsilon})\subset\mathcal{K}(\theta_*)\}$ for each $\theta_*\in\Theta_*$. 
Denote $\eta_*\triangleq\min_{\theta\in\mathbb{R}^{n_\theta}}\eta(\theta)$ the minimum value of $\eta$. Theorem \ref{thm: global convergence} below characterizes the optimality gap of FedGen. 
\begin{theorem}\label{thm: global convergence}
	{\em (Optimality gap).}
	Suppose $\Theta_*$ is non-empty,
and $\theta^{[i]}_0$ is independently uniformly sampled over a compact set $\Theta_0$ for all $i\in\mathcal{V}$, where
$\beta(\Theta_0\cap[\cup_{\theta_*\in\Theta_*}{\mathcal{B}(\theta_*,2\epsilon_0(\theta_*))}])>0$. Suppose all the conditions in Theorem \ref{thm: pareto improvement} hold. There exist $\omega\in(0,1]$ and  class $\mathcal{K}_\infty$ function $\kappa(\cdot)$ such that, $\forall i\in\mathcal{V}$ and any $\epsilon_1, \epsilon_2, \epsilon_3>0$, 
\begin{align}\label{ineq: optimality gap}
\eta(\theta^{[i]}_{\infty})-\eta_*\leqslant&\frac{L_\eta(q^{\max}+\epsilon_1)}{\alpha}+\epsilon_2+2\epsilon_3b^{\max}_\gamma
\end{align} 
 with probability at least
 \begin{align}\label{prob: for optimality gap}
 1-&\frac{(\sigma^{\max})^2}{\epsilon_1^2}-2\exp(-2\epsilon_2^2)-\frac{1}{\epsilon_3}-(1-\omega)^{|\mathcal{V}|}-\kappa(r^{\max}).\blacksquare
 \end{align}

\end{theorem}

\subsection{Discussion}\label{sec: discussion}

	{\em (Adjusting generalized safety through $b^{[i]}_\gamma$). } By (T2) in Theorem \ref{thm: monotonic optimality}, the probability of safe arrival in a new environment is lower bounded by  the (adjusted) empirical normalized arrival time $(1-\gamma)(1-y^{[i]}_k)$ and the estimation error term $(1-\gamma)b^{[i]}_\gamma$. Since $y^{[i]}_k\in[0,1]$, we always have $(1-\gamma)(1-y^{[i]}_k)\geqslant0$, the equality holds only when $y^{[i]}_k=1$, i.e., the policy $\pi_{\theta^{[i]}_k}$ renders collision in all the training environments and initial states. This also implies that $\gamma$ should be small in order to have a high safe arrival rate. Given any $\gamma\in(0,1)$, $b^{[i]}_\gamma$ in the error term $(1-\gamma)b^{[i]}_\gamma$ can be reduced to an arbitrarily small value by increasing $n^{[i]}_{\mathcal{E}}$ and $n_{int|\mathcal{E}}^{[i]}$ for any $\gamma>0$.
		
	{\em(Hyperparameter tuning of $r$ and $q^{[i]}$).}
	Similar to the literature in non-convex stochastic optimization \cite{fehrman2020convergence}\cite{ghadimi2013stochastic},  Theorem  \ref{thm: pareto improvement} requires hyperparameters $r$ and $q^{[i]}$ to satisfy certain conditions  that depend on parameters $L_{\nabla\eta}$ and $\sigma^{[i]}$, which can be unknown {\em a priori}. However, these parameters can be estimated numerically; e.g., $L_{\nabla\eta}$ can be estimated using finite differences and $\sigma^{[i]}$ can be estimated using empirical variance. In practice, 
	these conditions can also be satisfied by tuning $r$ small enough and $q^{[i]}$ large enough through trial and error,  a standard practice of hyperparameter tuning in training machine learning models, e.g., deep neural networks.

	{\em (Trade-off between consensus gap and improvement by the selection of $b^{[i]}_\gamma$).}
	Theorem \ref{thm: pareto improvement} (T4) implies that the consensus gap can be reduced  by reducing  $b^{[i]}_\gamma$ for all $i\in\mathcal{V}$. However, a small $b^{[i]}_\gamma$ can delay the convergence of the algorithm as Lemma \ref{lmm: finite switches} later shows that the number of times the learners adopts the estimates from the Cloud is upper bounded by $\frac{1}{\min_{j\in\mathcal{V}}b^{[j]}_\gamma}$.
	Similarly, there is also a trade-off in the selection of $b^{[i]}_\gamma$ in (T5) of  Theorem \ref{thm: pareto improvement}.
	Theorem \ref{thm: pareto improvement} (T5) shows that the improvement
	can be increased by increasing
	$b^{[i]}_\gamma$ for all $i\in\mathcal{V}$. However, as Lemma \ref{lmm: finite switches} later shows, this can reduce the number of times the learners adopt the estimates from the Cloud and hence reduce the probability $P\Big(\theta^{[i]}_\infty\neq\theta^{[i]}_{k^{[i]}_{fs}}\Big)$. This can eventually increase the total expectation $\mathbb{E}[\eta(\theta^{[i]}_\infty)-\eta(\theta^{[i]}_{k^{[i]}_{fs}})]$. Informally speaking, the selection of $b^{[i]}_\gamma$ determines the minimal gain learner $i$ demands after adopting the estimates from the Cloud. Therefore, larger $b^{[i]}_\gamma$ can prevent learner $i$ from adopting the estimates from the Cloud with small optimality improvement.
	Consider the  extreme case when $\min_{j\in\mathcal{V}}{b^{[j]}_\gamma}$ is so large that the learners would never adopt the estimates from the Cloud. Then we have $\theta^{[i]}_\infty=\theta^{[i]}_{k^{[i]}_{fs}}$ for all $i\in\mathcal{V}$, and there would be no improvement benefited from communication. 
	Nevertheless, the right hand side in (T5) of  Theorem \ref{thm: pareto improvement} is always non-positive, which implies that the adopted estimate is at least as optimal as the estimate without communication.
	
	{\em (The number of learners versus sample sizes in the learners).}
The upper bound in \eqref{ineq: optimality gap} implies that
smaller $q^{[j]}$ and smaller $b^{[j]}_\gamma$ for all $j\in\mathcal{V}$ can reduce the optimality gap. Recall the condition $q^{[j]}\geqslant4\sigma^{[j]}$ and the definition of $b^{[j]}_\gamma$ above Theorem \ref{thm: monotonic optimality}. Then  \eqref{ineq: optimality gap}  implies that large sample sizes, i.e., $n^{[j]}_{\mathcal{E}}$ and $n^{[j]}_{int|\mathcal{E}}$,  for all the learners can reduce the optimality gap. The probability bound \eqref{prob: for optimality gap} indicates that smaller variance of the estimation error $\sigma^{\max}$ and larger $|\mathcal{V}|$  can increase the probability of achieving the optimality gap in \eqref{ineq: optimality gap}. The class $\mathcal{K}_\infty$ function $\kappa(r^{\max})$ imposes a preference on small step size $r^{[j]}$.

\section{Proofs}\label{sec: proofs}

\subsection{Proof of Theorem \ref{thm: monotonic optimality}}
We first quantify the estimation error of $y_k^{[i]}$  and prove (T1). Then we  summarize the safety of the estimates and prove (T2).

The proof of 
(T1) is an adoption of Hoeffding's inequality below.
\begin{theorem}\label{thm: hoeffding}
	({\em Hoeffding's inequality, \cite{boucheron2013concentration}}). Let $q_1,\cdots, q_n$ be independent random variables such that $q_l$ takes its values in $[a_l,b_l]$ almost surely for all $1\leqslant l\leqslant n$. Then for every $\epsilon>0$, it holds that
	$$
	P\Big(|\sum_{l=1}^nq_l-\mathbb{E}[\sum_{l=1}^n q_l]|\geqslant \epsilon\Big)\leqslant 2\exp\Big(-\frac{2\epsilon^2}{\sum_{l=1}^n(b_l-a_l)^2}\Big). 
	$$ $\hfill\blacksquare$
\end{theorem}

\textbf{Proof of (T1):}
Assumptions \ref{assmp: stochastic environment} and \ref{assmp: stochastic initialization} imply $\mathbb{E}[J_{E}(x^{[i]}_{int|E};\pi_{\theta_k^{[i]}})]=\eta(\theta^{[i]}_k)$. 
Note that $J_E\in[0,1]$. 
Let $q_{ll'}\triangleq J_{E_l^{[i]}}(x^{[i]}_{int|E_l^{[i]},l'};\pi_{\theta_k^{[i]}})$ and hence $\mathbb{E}[q_{ll'}]=\mathbb{E}[J_{E_l^{[i]}}(x^{[i]}_{int|E_l^{[i]},l'};\pi_{\theta_k^{[i]}})]=\eta(\theta_k^{[i]})$.
 Then 
 \begin{align*}
&\sum_{l=1}^{n^{[i]}_{\mathcal{E}}}\sum_{l'=1}^{n^{[i]}_{int|\mathcal{E}}}q_{ll'}=\sum_{l=1}^{n^{[i]}_{\mathcal{E}}}\sum_{l'=1}^{n^{[i]}_{int|\mathcal{E}}}J_{E_l^{[i]}}(x^{[i]}_{int|E_l^{[i]},l'};\pi_{\theta_k^{[i]}})
 \\
 &\qquad\qquad\quad=n_{\mathcal{E}}^{[i]}n_{int|\mathcal{E}}^{[i]}y_k^{[i]},\\
 &\sum_{l=1}^{n^{[i]}_{\mathcal{E}}}\sum_{l'=1}^{n^{[i]}_{int|\mathcal{E}}}\mathbb{E}[q_{ll'}]=n^{[i]}_\mathcal{E}n^{[i]}_{int|\mathcal{E}}\eta(\theta_k^{[i]}).
 \end{align*}
Then Theorem \ref{thm: hoeffding} gives 
$
n_{\mathcal{E}}^{[i]}n_{int|\mathcal{E}}^{[i]}|y_k^{[i]}-\eta(\theta_k^{[i]})|\leqslant n_{\mathcal{E}}^{[i]}n_{int|\mathcal{E}}^{[i]}\epsilon
$ with probability at least $1-2\exp\Big(-2\epsilon^2n_{\mathcal{E}}^{[i]}n_{int|\mathcal{E}}^{[i]}\Big)$ for each $k\geqslant 0$. After some simple algebraic transformations, we have
\begin{align}\label{ineq: |y- eta|}
	|y_k^{[i]}-\eta(\theta_k^{[i]})|\leqslant\sqrt{\frac{\log (2/\gamma)}{2n_{\mathcal{E}}^{[i]}n_{int|\mathcal{E}}^{[i]}}},
\end{align}
 with probability at least $1-\gamma$, $\forall i\in\mathcal{V}$ and $k\geqslant 0$. 
$\hfill\blacksquare$

Notice that $J_{E}(x_{int};\pi_{\theta^{[i]}_k})\in[0,1]$ for any $E\in\mathcal{E}$ and $x_{int}\in\mathcal{X}$, and by definition of $J_E$, safe arrival is equivalent to $J_{E}(x_{int};\pi_{\theta^{[i]}_k})<1$.   Then the proof of (T2) is given as follows.

	\textbf{Proof of (T2): }
	(T1) renders that $
	\eta(\theta^{[i]}_k)\leqslant y^{[i]}_k+b^{[i]}_{\gamma}
	$
	with probability at least $1-\gamma$. Since Assumptions \ref{assmp: stochastic environment} and \ref{assmp: stochastic initialization} imply $\mathbb{E}[J_{E}(x_{int};\pi_{\theta^{[i]}_k})]=\eta(\theta^{[i]}_k)$, we have $\mathbb{E}[J_{E}(x_{int};\pi_{\theta^{[i]}_k})\mid \eta(\theta^{[i]}_k)\leqslant a]\leqslant a$ for any $a\in\mathbb{R}$. 
	Combining this with Markov's inequality (page 151, \cite{papoulis2002probability}), we have
	\begin{align*}
		&P\Big(J_{E}(x_{int};\pi_{\theta^{[i]}_k})\geqslant1\mid \eta(\theta^{[i]}_k)\leqslant y^{[i]}_k+b^{[i]}_{\gamma}\Big)\\
		&\leqslant \mathbb{E}[J_{E}(x_{int};\pi_{\theta^{[i]}_k})\mid \eta(\theta^{[i]}_k)\leqslant y^{[i]}_k+b^{[i]}_{\gamma}]\leqslant y^{[i]}_k+b^{[i]}_{\gamma}.
	\end{align*}
	Then we further have
	\begin{align}\label{ineq: P J eps}
		&P\Big(J_{E}(x_{int};\pi_{\theta^{[i]}_k})< 1, \eta(\theta^{[i]}_k)\leqslant y^{[i]}_k+b^{[i]}_{\gamma}\Big)\nonumber\\
		&=P\Big(J_{E}(x_{int};\pi_{\theta^{[i]}_k})< 1\mid \eta(\theta^{[i]}_k)\leqslant y^{[i]}_k+b^{[i]}_{\gamma}\Big)\nonumber\\
		&\quad \cdot P\Big( \eta(\theta^{[i]}_k)\leqslant y^{[i]}_k+b^{[i]}_{\gamma}\Big)\nonumber\\
		&\geqslant \big(1-(y^{[i]}_k+b^{[i]}_{\gamma})\big)(1-\gamma).
	\end{align}
	Notice that 
	\begin{align*}
	&P\Big(J_{E}(x_{int};\pi_{\theta^{[i]}_k})< 1\Big)\geqslant\\
	& P\Big(J_{E}(x_{int};\pi_{\theta^{[i]}_k})< 1, \eta(\theta^{[i]}_k)\leqslant y^{[i]}_k+b^{[i]}_{\gamma}\Big).
	\end{align*}
	Hence, the proof is concluded.
	$\hfill\blacksquare$





%
%

\subsection{Proof of Theorem \ref{thm: pareto improvement}}
In this section, we first provide a set of preliminary results in Section \ref{sec: preliminary results}, which mainly discusses the properties of the estimation of $z^{[i]}_{k-1}$ and  the estimates after the last time the learner adopts the estimate returned from the Cloud. Then the proofs of (T3), (T4) and (T5) of Theorem \ref{thm: pareto improvement} are presented in Sections \ref{proof: T3}, \ref{proof: T4} and \ref{proof: T5}, respectively.

To facilitate the proof, some important iterations of the algorithm FedGen are defined/repeated in Table \ref{table}.
	\begin{table}[h!]
	\centering 	 	
	\begin{tabular}{ | m{5em} | m{5cm}| } 
		\hline
		Symbol & Definition  \\ 
			\hline
		$k^{[i]}_{n}$, $n=1,2,\cdots$ & The iteration when  Lines \ref{ln: If eta < and xi >}-\ref{ln: converge false} are executed; i.e., learner $i$ adopts the estimates from the Cloud.\\ 
			\hline
		$k^{[i]}_*$&  The last time  Lines \ref{ln: If eta < and xi >}-\ref{ln: converge false} are executed. 
		If Lines \ref{ln: If eta < and xi >}-\ref{ln: converge false} are never executed,  then $k^{[i]}_*=0$.\\ 
		\hline
		$k^{[i]}_{fs}$&  The first time learner $i$ is stopped: $k^{[i]}_{fs}\triangleq\min\{k\geqslant0\mid \|z^{[i]}_{k}\|<q^{[i]}\}$.\\
		\hline
	$k^{[i]}_{ls}$& The last time learner $i$ is stopped: $k^{[i]}_{ls}\triangleq\min\{k\geqslant k^{[i]}_*\mid\|z^{[i]}_{k}\|<q^{[i]}\}$. \\
	
	\hline  
	
	\end{tabular}
	\caption{Definitions of important iterations}
	\label{table}
\end{table}

Notice that the above iterations satisfy:
\begin{align}\label{ineq: relation k}
k^{[i]}_{fs}+1\leqslant k^{[i]}_{1}< k^{[i]}_{2}<\cdots< k^{[i]}_{*}\leqslant k^{[i]}_{ls}.
\end{align}
 

%
%

\subsubsection{Preliminary results}\label{sec: preliminary results}

First of all, we provide the proof of Lemma \ref{lemma: unbiased estimator of z}.

{\bf Proof of Lemma \ref{lemma: unbiased estimator of z}: }
Assumption \ref{assmp: lipschitz} implies that $\eta$ is almost everywhere differentiable (Theorem 3.1.6 \cite{federer2014geometric}).
Hence,  Interchange of Differentiation and Integration (Corollary 2.8.7, \cite{bogachev2007measure}) and  Assumptions \ref{assmp: stochastic environment} and \ref{assmp: stochastic initialization} give
\begin{align*}
	&\mathbb{E}[z_k^{[i]}]=\mathbb{E}\Big[\nabla[ \frac{1}{n_{\mathcal{E}}^{[i]}n_{int|\mathcal{E}}^{[i]}}\sum_{l=1}^{n_{\mathcal{E}}^{[i]}}\sum_{l'=1}^{n_{int|\mathcal{E}}^{[i]}}J_{E^{[i]}_l}(x^{[i]}_{int|{E^{[i]}_l},l'};\pi_{\theta^{[i]}_k})]\Big]\nonumber\\
	&=\nabla\mathbb{E}[ \frac{1}{n_{\mathcal{E}}^{[i]}n_{int|\mathcal{E}}^{[i]}}\sum_{l=1}^{n_{\mathcal{E}}^{[i]}}\sum_{l'=1}^{n_{int|\mathcal{E}}^{[i]}}J_{E^{[i]}_l}(x^{[i]}_{int|{E^{[i]}_l},l'};\pi_{\theta^{[i]}_k})]\nonumber\\
	&=\nabla\eta(\theta^{[i]}_k).\qquad\qquad\qquad\qquad\qquad\qquad\qquad\qquad\qquad\hfill\blacksquare
\end{align*}

Denote the estimation error $\xi^{[i]}_{k}\triangleq \nabla\eta(\theta^{[i]}_{k})-z^{[i]}_{k}$. Lemma \ref{lemma: gradient estimate} quantifies   $\|\xi^{[i]}_{k}\|$.
\begin{lemma} \label{lemma: gradient estimate}
	Suppose Assumption \ref{assmp: ||z||} holds. Then it holds that $\|\xi^{[i]}_{k}\|\leqslant \epsilon$, $\epsilon>0$, with probability at least $1-\frac{(\sigma^{[i]})^2}{\epsilon^2}$. 
	
\end{lemma}	
\textbf{Proof:}
Combining Assumption \ref{assmp: ||z||} and Markov's inequality renders $\|\xi^{[i]}_{k}\|^2\geqslant \epsilon^2$, $\epsilon>0$, with probability at most $\frac{\mathbb{E}[\|\xi^{[i]}_{k}\|^2]}{\epsilon^2}\leqslant\frac{(\sigma^{[i]})^2}{\epsilon^2}$, or $\|\xi^{[i]}_{k}\|\leqslant \epsilon$ with probability at least $1-\frac{(\sigma^{[i]})^2}{\epsilon^2}$.
$\hfill\blacksquare$

The following lemma  provides a property of the expectation of $\|\xi^{[i]}_{k}\|$. 
\begin{lemma}\label{lmm: E u}
	It holds that $\mathbb{E}[\|\xi^{[i]}_{k}\|]=\int_{0}^\infty P\Big(\|\xi^{[i]}_{k}\|> t\Big)dt$.  	
	
	{\bf Proof:} 
 For all $t\geqslant0$, it holds that
$t(1-P\Big(\|\xi^{[i]}_{k}\|\leqslant t\Big))\geqslant0.
$
By Lemma \ref{lemma: gradient estimate}, we also have $$
\lim_{t\to\infty}t(1-P\Big(\|\xi^{[i]}_{k}\|\leqslant t\Big))\leqslant\lim_{t\to\infty} t(1-(1-\frac{(\sigma^{[i]})^2}{t^2}))=0.
$$ Therefore, $\lim_{t\to\infty}t(1-P\Big(\|\xi^{[i]}_{k}\|\leqslant t\Big))=0$.
	Denote $p(\cdot)$ the probability density function of random variable $\|\xi^{[i]}_{k}\|$. 	
	By integration by parts, we have
	\begin{align*}
		&\int_{0}^\infty (1-P\Big(\|\xi^{[i]}_{k}\|\leqslant t\Big))dt=t(1-P\Big(\|\xi^{[i]}_{k}\|\leqslant t\Big))\Big|_{t=0}^\infty\\
		&\qquad\quad+\int_{0}^\infty tp(\|\xi^{[i]}_{k}\|=t)dt=\int_{0}^\infty tp(\|\xi^{[i]}_{k}\|=t)dt.
	\end{align*}
Since $\|\xi^{[i]}_{k}\|\geqslant 0$, we have $p(\|\xi^{[i]}_{k}\|=t)=0$ for all $t<0$. Therefore, we have
\begin{align*}
\mathbb{E}[\|\xi^{[i]}_{k}\|]&=\int_{-\infty}^\infty tp(\|\xi^{[i]}_{k}\|=t)dt=\int_{0}^\infty tp(\|\xi^{[i]}_{k}\|=t)dt\\
&=\int_{0}^\infty P\Big(\|\xi^{[i]}_{k}\|> t\Big)dt.\qquad\qquad\qquad\qquad\quad\blacksquare
\end{align*}

\end{lemma}

The following lemma finds a lower bound of $\langle \nabla \eta(\theta^{[i]}_{k-1}-\lambda z^{[i]}_{k-1}),z^{[i]}_{k-1}\rangle$ for all $\lambda\in[0,\frac{r^{[i]}}{k^\rho}]$.
\begin{lemma}\label{lemma: dot nabla hat theta z k-1}
	Suppose Assumptions \ref{assmp: smooth}  and  \ref{assmp: ||z||} hold. It holds that, for any $\epsilon>0$ and $\lambda\in[0,\frac{r^{[i]}}{k^\rho}]$, 
	\begin{align*}
	\langle \nabla \eta(\theta^{[i]}_{k-1}-\lambda z^{[i]}_{k-1}),z^{[i]}_{k-1}\rangle\geqslant&(1-L_{\nabla\eta}\frac{r^{[i]}}{k^\rho})\|z^{[i]}_{k-1}\|^2\\
	&-\|\xi^{[i]}_{k-1}\|\|z^{[i]}_{k-1}\|.
	\end{align*}
\end{lemma}	
{\bf Proof:}
Denote $\nu\triangleq \nabla \eta(\theta^{[i]}_{k-1})-  \nabla \eta(\theta^{[i]}_{k-1}-\lambda z^{[i]}_{k-1})$. 
Write
\begin{align}
	&\langle  \nabla \eta(\theta^{[i]}_{k-1}-\lambda z^{[i]}_{k-1}),z^{[i]}_{k-1}\rangle= \langle \nabla\eta(\theta^{[i]}_{k-1})-\nu,z^{[i]}_{k-1}\rangle\nonumber\\
	&= \langle \nabla\eta(\theta^{[i]}_{k-1}),z^{[i]}_{k-1}\rangle-\langle\nu,z^{[i]}_{k-1}\rangle.\label{ineq: dot product nabla eta hat theta}
\end{align}
Next we find the lower bounds of the two terms on the right hand side of \eqref{ineq: dot product nabla eta hat theta}.
Consider the first term.  
Then we have
\begin{align}
	\langle \nabla\eta(\theta^{[i]}_{k-1}),z^{[i]}_{k-1}\rangle&=\langle z^{[i]}_{k-1}+\xi^{[i]}_{k-1},z^{[i]}_{k-1}\rangle\nonumber\\
	&=\|z^{[i]}_{k-1}\|^2+\langle \xi^{[i]}_{k-1},z^{[i]}_{k-1}\rangle.\label{eq: dot nabla eta z}
\end{align}
By the Cauchy-Schwartz inequality, we have
\begin{align}\label{ineq: dot nabla eta z}
	\langle \nabla\eta(\theta^{[i]}_{k-1}),z^{[i]}_{k-1}\rangle\geqslant \|z^{[i]}_{k-1}\|^2-\|\xi^{[i]}_{k-1}\|\|z^{[i]}_{k-1}\|.
\end{align}

Consider the second term in \eqref{ineq: dot product nabla eta hat theta}.
Assumption \ref{assmp: smooth} implies
\begin{align}\label{ineq: ||nu||}
	\|\nu\|&\leqslant L_{\nabla\eta} \|\theta^{[i]}_{k-1}-(\theta^{[i]}_{k-1}-\lambda z^{[i]}_{k-1})\|=L_{\nabla\eta}\lambda\|z^{[i]}_{k-1}\|\nonumber\\
	&\leqslant L_{\nabla\eta}\frac{r^{[i]}}{k^\rho}\|z^{[i]}_{k-1}\|.
\end{align}
Using the Cauchy-Schwartz inequality and \eqref{ineq: ||nu||} render
\begin{align}\label{ineq: dot nu z}
	\langle\nu,z^{[i]}_{k-1}\rangle\leqslant \|\nu\|\|z^{[i]}_{k-1}\|\leqslant L_{\nabla\eta}\frac{r^{[i]}}{k^\rho}\|z^{[i]}_{k-1}\|^2.
\end{align}

Combining \eqref{ineq: dot nabla eta z} and \eqref{ineq: dot nu z}  with \eqref{ineq: dot product nabla eta hat theta} gives
the lemma.
$\hfill\blacksquare$

Next Lemma \ref{lmm: finite switches} shows that each learner $i$ only adopts the estimates from the Cloud for a finite number of times. 
\begin{lemma}\label{lmm: finite switches}
	It holds that  $n\leqslant\frac{1}{\min_{j\in\mathcal{V}}{b^{[j]}_\gamma}}$ for all $k^{[i]}_{n}$, $i\in\mathcal{V}$.
\end{lemma}	
\textbf{Proof:}
Pick any $i\in\mathcal{V}$.            
Note that when  Lines \ref{ln: If eta < and xi >}-\ref{ln: converge false} are executed at iteration $k^{[i]}_{n}$, we must have 
\begin{align}\label{eq: zeta decrease}
	\zeta^{[i]}_{k^{[i]}_{n}}=y^{[j]}_{l}<\zeta^{[i]}_{k^{[i]}_{n}-1}-b^{[j]}_\gamma\leqslant \zeta^{[i]}_{k^{[i]}_{n}-1}-b^{\min}_\gamma,
\end{align}
where $(j,l)= \arg\min_{i\in\mathcal{V}, l'=0,\cdots, k^{[i]}_{n}-1}y^{[i]}_{l'}+b^{[i]}_\gamma$.
Since initialization gives $\zeta^{[i]}_{0}=1$, \eqref{eq: zeta decrease} implies
\begin{align}\label{ineq: zeta ineq}
	\zeta^{[i]}_{k^{[i]}_{n}}\leqslant 1- nb^{\min}_\gamma.    
\end{align}
Since $\zeta^{[i]}_{k^{[i]}_{n}}\in[0,1]$, \eqref{ineq: zeta ineq} renders $n\leqslant\frac{1}{b^{\min}_\gamma}$. 
$\hfill\blacksquare$

Next we show that the event $\|z^{[i]}_{k}\|<q^{[i]}$ happens almost surely, which indicates convergence to a local minimum, by showing the almost sure existence of $k^{[i]}_{ls}$.
\begin{lemma}\label{lmm: existence of k}
	Suppose Assumptions  \ref{assmp: stochastic environment}, \ref{assmp: stochastic initialization},  \ref{assmp: lipschitz}, \ref{assmp: smooth} and \ref{assmp: ||z||} hold. If $q^{[i]}\geqslant 4\sigma^{[i]}$, then it holds that $k^{[i]}_{ls}$ exists almost surely. 
	
	{\bf Proof:} 
	By definition of $k^{[i]}_{ls}$, we have $\|z^{[i]}_{k}\|\geqslant q^{[i]}$ for all $k\in{[k^{[i]}_*,k^{[i]}_{ls}]}$ and hence Lines \ref{ln: If eta < and xi >}-\ref{ln: converge false} are never executed for all $k\in{[k^{[i]}_*,k^{[i]}_{ls}]}$. Denote event $A\triangleq\{k^{[i]}_{ls} \textrm{ exists.}\}$ and the complement 	$A^c\triangleq\{k^{[i]}_{ls} \textrm{ does not exist.}\}$. Notice that we can equivalently write $A^c=\{\|z^{[i]}_k\|\geqslant q^{[i]}, \forall k\geqslant k^{[i]}_*\}$. Then $A^c$ implies Lines \ref{ln: Else} and \ref{ln: proj SV2} are executed for all $k\geqslant k^{[i]}_*$ and hence $\theta^{[i]}_k=\hat{\theta}^{[i]}_k=\theta^{[i]}_{k-1}-\frac{r^{[i]}}{k^\rho} z^{[i]}_{k-1}$ for all $k\geqslant k^{[i]}_*$, which is a  stochastic gradient descent step \cite{ghadimi2013stochastic}. 	Given Assumptions \ref{assmp: lipschitz}, \ref{assmp: smooth} and \ref{assmp: ||z||},  and  Lemma \ref{lemma: unbiased estimator of z}, Corollary 3.3 and inequality (3.32) in \cite{ghadimi2013stochastic} show that $\|\nabla\eta(\theta^{[i]}_k)\|\to 0$ almost surely. Then, for any $\delta>0$, there exists some $K_\delta>k^{[i]}_*$ such that $\|\nabla\eta(\theta^{[i]}_k)\|< \delta$ for all $k\geqslant K_\delta$ almost surely. Since $q^{[i]}\geqslant4\sigma^{[i]}$, we can pick $\delta\in(0,\sigma^{[i]})$ and let $\epsilon\triangleq q^{[i]}-\delta$. By the above construction, we have $\epsilon>\sigma^{[i]}$.
	Then Lemma \ref{lemma: gradient estimate}  implies 
	\begin{align}\label{ineq: ||z||}
		\|z_{k}^{[i]}\|=&\|z_{k}^{[i]}-\nabla\eta(\theta_{k}^{[i]})+\nabla\eta(\theta_{k}^{[i]})\|\leqslant
		\|z_{k}^{[i]}-\nabla\eta(\theta_{k}^{[i]})\|\nonumber\\
		&+\|\nabla\eta(\theta_{k}^{[i]})\|\leqslant \epsilon+\|\nabla\eta(\theta_{k}^{[i]})\|<q^{[i]}
	\end{align}
	with probability at least $ 1-\frac{(\sigma^{[i]})^2}{\epsilon^2}$, $\frac{(\sigma^{[i]})^2}{\epsilon^2}<1$, for each $k\geqslant K_\delta$. 
	Due to the independent estimate of $z^{[i]}_k$ over $k$, we have
	\begin{align*}
		P\Big(A^c\Big)&=
		\lim_{\tilde{k}\to\infty}P\Big( \|z_{k}^{[i]}\|\geqslant q^{[i]},\forall k\in [k^{[i]}_*,\tilde{k}]\Big)\\
		&\leqslant\lim_{\tilde{k}\to\infty}P\Big( \|z_{k}^{[i]}\|\geqslant q^{[i]},\forall k\in [K_\delta,\tilde{k}]\Big)\\
		&\leqslant \lim_{\tilde{k}\to\infty} \Big(\frac{(\sigma^{[i]})^2}{\epsilon^2}\Big)^{\tilde{k}-K_\delta}=0.
	\end{align*}
	Therefore, $P\Big(A\Big)=1-P\Big(A^c\Big)=1$. 
	$\hfill\blacksquare$
\end{lemma}	

The following lemma shows that  $\eta(\theta^{[i]}_{k})\leqslant\eta(\theta^{[i]}_{k^{[i]}_*})$, for all $k\geqslant k^{[i]}_*$ in expectation.
\begin{lemma}\label{lmm: eta theta > theta *}
	Suppose Assumptions \ref{assmp: smooth}  and  \ref{assmp: ||z||} hold,  $r^{[i]}\leqslant \frac{1}{2L_{\nabla\eta}}$ and $q^{[i]}\geqslant 4\sigma^{[i]}$.  It holds that
		$
		\mathbb{E}[\eta(\theta^{[i]}_k)-\eta(\theta^{[i]}_{k^{[i]}_*})]\leqslant0
		$ for all $k\geqslant k^{[i]}_*$. 
	
	{\bf Proof:}  Recall that $k^{[i]}_*$ is the last time learner $i$ adopts the estimate from the Cloud,  and Lemma \ref{lmm: finite switches} shows that $k^{[i]}_*$ exists. 
	Note that Figure \ref{fig: logic} indicates that $\theta^{[i]}_{k}=\theta^{[i]}_{k^{[i]}_{ls}}=\theta^{[i]}_\infty$ 	for all $k\geqslant k^{[i]}_{ls}$. When $k^{[i]}_{ls}=k^{[i]}_*$,  we have	$
	\mathbb{E}[\eta(\theta^{[i]}_k)-\eta(\theta^{[i]}_{k^{[i]}_*})]=0
	$ for all $k\geqslant k^{[i]}_*$.
	Hence,  in the sequel, we consider the case where  $
	\theta^{[i]}_k=\hat{\theta}^{[i]}_k =\theta^{[i]}_{k-1}-\frac{r^{[i]}}{k^\rho} z^{[i]}_{k-1}   
	$ is executed for all $k\in[k^{[i]}_*+1,k^{[i]}_{ls}]$, when $k^{[i]}_{ls}\geqslant k^{[i]}_*+1$.
	
	Denote $g:\mathbb{R}\to\mathbb{R}$ such that $g(\lambda)\triangleq \eta(\theta^{[i]}_{k-1}-\lambda z^{[i]}_{k-1})$. Then by chain rule, we have 
	$$
	\frac{d}{d\lambda}g(\lambda)= -\langle \nabla \eta(\theta^{[i]}_{k-1}-\lambda z^{[i]}_{k-1}),z^{[i]}_{k-1}\rangle. 
	$$
	Therefore, we have
	\begin{align*}
		&\eta(\theta^{[i]}_{k})-\eta(\theta^{[i]}_{k-1})=\eta(\theta^{[i]}_{k-1}-\frac{r^{[i]}}{k^\rho} z^{[i]}_{k-1})-\eta(\theta^{[i]}_{k-1})\\
		&=g(\frac{r^{[i]}}{k^\rho})-g(0)=\int_{0}^{\frac{r^{[i]}}{k^\rho} }\frac{d}{d\lambda}g(\lambda)d\lambda\\
		&=-\int_{0}^{\frac{r^{[i]}}{k^\rho} }\langle \nabla \eta(\theta^{[i]}_{k-1}-\lambda z^{[i]}_{k-1}),z^{[i]}_{k-1}\rangle d\lambda.
	\end{align*}
	Combining this with Lemma \ref{lemma: dot nabla hat theta z k-1}, we have
	\begin{align*}
		\eta(\theta^{[i]}_{k})-\eta(\theta^{[i]}_{k-1})\leqslant&-\frac{r^{[i]}}{k^\rho}\big((1-L_{\nabla\eta}\frac{r^{[i]}}{k^\rho})\|z^{[i]}_{k-1}\|^2\\
		&-\|\xi^{[i]}_{k-1}\|\|z^{[i]}_{k-1}\|\big).
	\end{align*}
	For notational simplicity, we denote
	\begin{align*}
	\delta^{[i]}_{k}&\triangleq \eta(\theta^{[i]}_{k})-\eta(\theta^{[i]}_{k-1}),~	b^{[i]}_{k-1}\triangleq\frac{r^{[i]}}{k^\rho} \|z^{[i]}_{k-1}\| \\
	a^{[i]}_{k-1}&\triangleq \frac{r^{[i]}}{k^\rho}(1-L_{\nabla\eta}\frac{r^{[i]}}{k^\rho})\|z^{[i]}_{k-1}\|^2.
	\end{align*}
 Therefore, the above inequality can be rewritten to
	\begin{align}\label{ineq: P u<= a bc}
		\delta^{[i]}_{k}\leqslant -a^{[i]}_{k-1}+\|\xi^{[i]}_{k-1}\| b^{[i]}_{k-1}.
	\end{align}

Combining Lemma \ref{lmm: E u} and Markov's inequality renders
		\begin{align*}
			\mathbb{E}[\|\xi^{[i]}_{k}\|]&=\int_{0}^{\sigma^{[i]}}P\Big(\|\xi^{[i]}_k\|> t\Big)dt+\int^{\infty}_{\sigma^{[i]}}P\Big(\|\xi^{[i]}_k\|> t\Big)dt\\
			&\leqslant \sigma^{[i]}+\int^{\infty}_{\sigma^{[i]}}\frac{(\sigma^{[i]})^2}{t^2}dt=2\sigma^{[i]}.		
		\end{align*} 
		for all $k\geqslant 1$.	
		Therefore, combining this with \eqref{ineq: P u<= a bc} implies
		\begin{align}
			\mathbb{E}[\delta^{[i]}_{k}\mid z^{[i]}_{k-1}]
			&\leqslant\mathbb{E}[-a^{[i]}_{k-1}+\|\xi^{[i]}_{k-1}\| b^{[i]}_{k-1}\mid z^{[i]}_{k-1}]\nonumber\\
			&= -a^{[i]}_{k-1}+b^{[i]}_{k-1}\mathbb{E}[\|\xi^{[i]}_{k-1}\|]\nonumber\\
			&\leqslant-a^{[i]}_{k-1}+2b^{[i]}_{k-1}\sigma^{[i]}.\label{ineq: E theta |theta}
		\end{align}

		Since $k\in[k^{[i]}_*+1,k^{[i]}_{ls}]$, $\|z^{[i]}_{k-1}\|\geqslant q^{[i]}$. Plugging in the definitions of $a^{[i]}_{k-1}$ and $b^{[i]}_{k-1}$ and combining with $r^{[i]}\leqslant \frac{1}{2L_{\nabla\eta}}$ renders
		\begin{align}\label{ineq: a / b geq q}
			\frac{a^{[i]}_{k-1}}{b^{[i]}_{k-1}}=(1-L_{\nabla\eta} r^{[i]}/k^\rho)\|z^{[i]}_{k-1}\|&\geqslant\frac{(q^{[i]})}{2}.
		\end{align}

		Since $q^{[i]}> 4\sigma^{[i]}$, \eqref{ineq: a / b geq q} renders that $\frac{a^{[i]}_{k-1}}{b^{[i]}_{k-1}}\geqslant 2\sigma^{[i]}$ and hence $-a^{[i]}_{k-1}+2b^{[i]}_{k-1}\sigma^{[i]}\leqslant 0$ for $k\in[k^{[i]}_*+1,k^{[i]}_{ls}]$. Then combining this with \eqref{ineq: E theta |theta} renders $\mathbb{E}[\delta^{[i]}_{k}\mid z^{[i]}_{k-1}]\leqslant 0$, which implies
		\begin{align}\label{ineq: delta <0}
			\mathbb{E}[\delta^{[i]}_{k}]=\int\mathbb{E}[\delta^{[i]}_{k}\mid z^{[i]}_{k-1}] p(z^{[i]}_{k-1}) dz^{[i]}_{k-1}\leqslant 0,
		\end{align}
		for all $k\in[k^{[i]}_*+1,k^{[i]}_{ls}]$.

		Notice that the definition of $\delta^{[i]}_{k}$ renders 
		$$
		\eta^{[i]}(\theta^{[i]}_{k})-\eta^{[i]}(\theta^{[i]}_{k^{[i]}_*})=\sum_{k'=k^{[i]}_*+1}^{k}{\delta^{[i]}_{k'}},
		$$
		for any $k\geqslant k^{[i]}_*+1$.
		Then by \eqref{ineq: delta <0} we have 
		\begin{align*}	 		&\mathbb{E}[\eta^{[i]}(\theta^{[i]}_{k})-\eta^{[i]}(\theta^{[i]}_{k^{[i]}_*})]=\mathbb{E}[\sum_{k'=k^{[i]}_*+1}^{k}{\delta^{[i]}_{k'}}]\\
			&=\sum_{k'=k^{[i]}_*+1}^{k}\mathbb{E}[\delta^{[i]}_{k'}]\leqslant 0.
		\end{align*}
		The proof is conluded. $\hfill\blacksquare$
\end{lemma}

\subsubsection{Proof of (T3) in Theorem \ref{thm: pareto improvement}}\label{proof: T3}
Lemma \ref{lmm: existence of k} shows that $k^{[i]}_{ls}$ exists almost surely. Therefore,  Lines \ref{ln: proj SV2} and \ref{ln: Else} implies that $\theta^{[i]}_k=\hat{\theta}^{[i]}_k=\theta^{[i]}_{k-1}$ for all $k\geqslant k^{[i]}_{ls}+1$ and hence $\lim_{k\to\infty}\theta^{[i]}_k=\theta^{[i]}_\infty=\theta^{[i]}_{k^{[i]}_{ls}}$. $\hfill\blacksquare$

\subsubsection{Proof of (T4) in Theorem \ref{thm: pareto improvement}}\label{proof: T4}


Notice that for any $k,k'\geqslant1$ it holds that
\begin{align*}
&\mathbb{E}[\eta(\theta^{[i]}_k)-\eta(\theta^{[j]}_{k'})]\\
&=\mathbb{E}[\eta(\theta^{[i]}_k)-y^{[i]}_k+y^{[i]}_k-\eta(\theta^{[j]}_{k'})-y^{[j]}_{k'}+y^{[j]}_{k'}].
\end{align*}
Since estimation error $\eta(\theta^{[i]}_k)-y^{[i]}_{k}$ is independent of $\theta^{[i]}_{k}$ and Assumptions \ref{assmp: stochastic environment} and \ref{assmp: stochastic initialization} imply $\mathbb{E}[\eta(\theta^{[i]}_k)-y^{[i]}_k]=0$, the above equality becomes
\begin{align}\label{eq: E eta - eta = E y - y}
	&\mathbb{E}[\eta(\theta^{[i]}_k)-\eta(\theta^{[j]}_{k'})]=\mathbb{E}[y^{[i]}_k-y^{[j]}_{k'}].
\end{align}


Recall that  Lemma \ref{lmm: existence of k} shows that $\theta^{[i]}_{k^{[i]}_{ls}}$ exists almost surely.
Denote $j^*\triangleq\arg\min_{j\in\mathcal{V}}\eta(\theta^{[j]}_{{k^{[j]}_{ls}}})$.
Since learner $i$ does not execute Line \ref{ln: If eta < and xi >} at iteration $k^{[i]}_{ls}$, we have
$$y^{[j^*]}_{k^{[j^*]}_{ls}}+b^{[j^*]}_\gamma\geqslant\min\{y^{[i]}_{k^{[i]}_{ls}}-b^{[i]}_\gamma, \zeta^{[i]}_{k^{[i]}_{ls}}\}.$$  We now distinguish two cases.

Case 1: $y^{[i]}_{k^{[i]}_{ls}}-b^{[i]}_\gamma<\zeta^{[i]}_{k^{[i]}_{ls}}$. This implies  $y^{[j^*]}_{k^{[j^*]}_{ls}}+b^{[j^*]}_\gamma\geqslant y^{[i]}_{k^{[i]}_{ls}}-b^{[i]}_\gamma$, or 
\begin{align}\label{ineq: E ykc- ykc}
y^{[i]}_{k^{[i]}_{ls}}-y^{[j^*]}_{k^{[j^*]}_{ls}}\leqslant b^{[i]}_\gamma+b^{[j^*]}_\gamma\leqslant2\max_{j\in\mathcal{V}}b^{[j]}_\gamma.
\end{align}


Case 2: $\zeta^{[i]}_{k^{[i]}_{ls}}\leqslant y^{[i]}_{k^{[i]}_{ls}}-b^{[i]}_\gamma$. Line \ref{ln: zeta = zeta -} implies 
\begin{align*}
&y^{[j^*]}_{k^{[j^*]}_{ls}}+b^{[j^*]}_\gamma\geqslant \zeta^{[i]}_{k^{[i]}_{ls}}=\zeta^{[i]}_{k^{[i]}_*}=y^{[j]}_{l},\\
&(j,l)=\arg\min_{i\in\mathcal{V}, l'=0,\cdots, k^{[i]}_*-1}y^{[i]}_{l'}+b^{[i]}_\gamma
\end{align*}
Therefore, $y^{[j]}_{l}-y^{[j^*]}_{k^{[j^*]}_{ls}}\leqslant b^{[j^*]}_\gamma$. Recall that  Line \ref{ln: proj SV1} implies 
$\theta^{[i]}_{k^{[i]}_*}=\theta^{[j]}_l$ and hence $y^{[i]}_{k^{[i]}_*}=y^{[j]}_{l}$.
This renders
\begin{align}\label{ineq: theta k* - theta j *}
	y^{[i]}_{k^{[i]}_*}-y^{[j^*]}_{k^{[j^*]}_{ls}}\leqslant b^{\max}_\gamma.
\end{align}
Lemma \ref{lmm: eta theta > theta *} and \eqref{eq: E eta - eta = E y - y} render  $\mathbb{E}[y^{[i]}_{k^{[i]}_{ls}}-y^{[i]}_{k^{[i]}_*}]=	\mathbb{E}[\eta(\theta^{[i]}_{k^{[i]}_{ls}})-\eta(\theta^{[i]}_{k^{[i]}_*})]\leqslant 0.$ Combining this with \eqref{ineq: theta k* - theta j *} renders
\begin{align}\label{ineq: E ykc- ykc case 2}
	&\mathbb{E}[y^{[i]}_{k^{[i]}_{ls}}-y^{[j^*]}_{k^{[j^*]}_{ls}}]
	=\mathbb{E}[y^{[i]}_{k^{[i]}_{ls}}-y^{[i]}_{k^{[i]}_*}+y^{[i]}_{k^{[i]}_*}-y^{[j^*]}_{k^{[j^*]}_{ls}}]\nonumber\\
	&=\mathbb{E}[y^{[i]}_{k^{[i]}_{ls}}-y^{[i]}_{k^{[i]}_*}]+\mathbb{E}[y^{[i]}_{k^{[i]}_*}-y^{[j^*]}_{k^{[j^*]}_{ls}}]	\leqslant\mathbb{E}[y^{[i]}_{k^{[i]}_*}-y^{[j^*]}_{k^{[j^*]}_{ls}}]\nonumber\\
	&\leqslant b^{\max}_\gamma.
\end{align}
By \eqref{eq: E eta - eta = E y - y}, combining  \eqref{ineq: E ykc- ykc} and \eqref{ineq: E ykc- ykc case 2} renders
\begin{align*}
	&\mathbb{E}[\eta(\theta^{[i]}_{k^{[i]}_{ls}})-\eta(\theta^{[j^*]}_{k^{[j^*]}_{ls}})]=\mathbb{E}[y^{[i]}_{k^{[i]}_{ls}}-y^{[j^*]}_{k^{[j^*]}_{ls}}]\leqslant 2b^{\max}_\gamma.
\end{align*}
 Recall that $k^{[i]}_*$  is the last time adopting estimates from the Cloud (Lines \ref{ln: If eta < and xi >}-\ref{ln: converge false} are executed).
	Figure \ref{fig: logic} implies that $\theta^{[i]}_k=\hat{\theta}^{[i]}_k=\theta^{[i]}_{k-1}$ for all $k\geqslant k^{[i]}_{ls}+1$ and hence $\lim_{k\to\infty}\theta^{[i]}_k=\theta^{[i]}_\infty=\theta^{[i]}_{k^{[i]}_{ls}}$. Therefore, we have $\theta^{[i]}_{\infty}=\theta^{[i]}_{k^{[i]}_{ls}}$ for all $i\in\mathcal{V}$. Hence, the above inequality implies that,  for any $i\in\mathcal{V}$, $$\mathbb{E}[\eta(\theta^{[i]}_\infty)-\eta(\theta^{[j^*]}_\infty)]=\mathbb{E}[y^{[i]}_\infty-y^{[j^*]}_\infty]\leqslant 2b^{\max}_\gamma.\qquad\blacksquare$$

\subsubsection{Proof of (T5) in Theorem \ref{thm: pareto improvement}}\label{proof: T5}

Since  Lemma \ref{lmm: existence of k}  shows that $k^{[i]}_{ls}$ exists almost surely, by \eqref{ineq: relation k}, we have $k^{[i]}_{fs}$ exists almost surely. 
 Recall that $k^{[i]}_{fs}+1\leqslant k^{[i]}_{1}$ from \eqref{ineq: relation k}.
 Notice that at iteration $k^{[i]}_{fs}$, agent $i$ stops its local gradient descent, and its estimate remains the same for the following iterations until it adopts an estimate from the Cloud.
 Since $\theta^{[i]}_\infty\neq \theta^{[i]}_{k^{[i]}_{fs}}$, agent $i$ adopts estimates from the Cloud, executing  Lines \ref{ln: If eta < and xi >}-\ref{ln: converge false}, at least once after iteration $k^{[i]}_{fs}$. This implies that $k^{[i]}_*\geqslant k^{[i]}_1\geqslant 1$ and
\begin{align}\label{eq: theta=theta}
\theta^{[i]}_{k^{[i]}_{fs}}=\theta^{[i]}_{k^{[i]}_{fs}+1}=\cdots=\theta^{[i]}_{k^{[i]}_1-1}.
\end{align}

%
%
	Recall that (T3) of Theorem \ref{thm: pareto improvement} shows that $\theta^{[i]}_\infty$ exists almost surely.
	By Lemma \ref{lmm: finite switches}, $k^{[i]}_*$ exists.
	Since $k^{[i]}_*\geqslant k^{[i]}_1\geqslant 1$, Lines \ref{ln: If eta < and xi >}-\ref{ln: converge false}  imply that there exists
	$(j_1,l_1)= \arg\min_{i\in\mathcal{V}, l'=0,\cdots, k^{[i]}_{1}-1}\allowbreak y^{[i]}_{l'}+b^{[i]}_\gamma$ such that $y^{[j_1]}_{l_1}+b^{[j_1]}_\gamma<y^{[i]}_{{k^{[i]}_1}-1}-b^{[i]}_\gamma$. Consider $(j_*,l_*)= \arg\min_{i\in\mathcal{V}, l'=0,\cdots, k^{[i]}_{*}-1}\allowbreak y^{[i]}_{l'}+b^{[i]}_\gamma$. It is obvious that $y^{[j_*]}_{l_*}+b^{[j_*]}_\gamma\leqslant y^{[j_1]}_{l_1}+b^{[j_1]}_\gamma<y^{[i]}_{{k^{[i]}_1}-1}-b^{[i]}_\gamma$, or 
	\begin{align}\label{ineq: y-y}
	y^{[j^*]}_{l_*}-y^{[i]}_{{k^{[i]}_1}-1}<-(b^{[i]}_\gamma+b^{[j_*]}_\gamma).
	\end{align}
		Since learner $i$ adopts the estimate from the Cloud, i.e., executes Lines \ref{ln: If eta < and xi >}-\ref{ln: converge false}, at iteration $k^{[i]}_{*}$, Line \ref{ln: proj SV1} implies $\theta^{[i]}_{k^{[i]}_{*}}=\theta^{[j_*]}_{l_*}$.
	Following the same logic of \eqref{eq: E eta - eta = E y - y} and combining with \eqref{ineq: y-y}, we have
	\begin{align*}
		&\mathbb{E}[\eta(\theta^{[i]}_{k^{[i]}_{*}})-\eta(\theta^{[i]}_{{k^{[i]}_1}-1})]\\
		&=\mathbb{E}[\eta(\theta^{[i]}_{k^{[i]}_{*}})-y^{[j^*]}_{l_*}+y^{[j^*]}_{l_*}-\eta(\theta^{[i]}_{{k^{[i]}_1}-1})+y^{[i]}_{{k^{[i]}_1}-1}-y^{[i]}_{{k^{[i]}_1}-1}]\\
	&=\mathbb{E}[y^{[j^*]}_{l_*}-y^{[i]}_{{k^{[i]}_1}-1}]< -(b^{[i]}_\gamma+b^{[j_*]}_\gamma).
	\end{align*}
Combining this with Lemma \ref{lmm: eta theta > theta *} renders
\begin{align*}
&\mathbb{E}[\eta(\theta^{[i]}_\infty)- \eta(\theta^{[i]}_{k^{[i]}_1-1})]\\
&=\mathbb{E}[\eta(\theta^{[i]}_\infty)-\eta(\theta^{[i]}_{k^{[i]}_{*}})+\eta(\theta^{[i]}_{k^{[i]}_{*}})- \eta(\theta^{[i]}_{k^{[i]}_1-1})]\\
&=\mathbb{E}[\eta(\theta^{[i]}_{\infty})-\eta(\theta^{[i]}_{k^{[i]}_{*}})]+\mathbb{E}[ \eta(\theta^{[i]}_{k^{[i]}_{*}})-\eta(\theta^{[i]}_{k^{[i]}_1-1})]\\
&<-(b^{[i]}_\gamma+b^{[j_*]}_\gamma)\leqslant -2b^{\min}_\gamma.
\end{align*}
Combining this with $\theta^{[i]}_{k^{[i]}_{fs}}=\theta^{[i]}_{k^{[i]}_1-1}$ in \eqref{eq: theta=theta}, the proof is concluded. $\hfill\blacksquare$

\subsection{Proof of Theorem \ref{thm: global convergence}}\label{proof: theorem global convergence}
For notational simplicity, we define two closed neighborhoods for each $\theta_*\in\Theta_*$: $\Psi(\theta_*)\triangleq \mathcal{B}(\theta_*,4\epsilon_0(\theta_*)+2\sqrt{\epsilon_0(\theta_*)})$ and $\Psi_1(\theta_*)\triangleq\mathcal{B}(\theta_*,2\epsilon_0(\theta_*))$.
Then the proof of the theorem is composed of four parts. First, we assume that there exists some $i\in\mathcal{V}$ such that $\theta^{[i]}_{k^{[i]}_{fs}}\in\Psi(\theta_*)$ for some $\theta_*\in\Theta_*$ and derive the probabilistic upper bound of $\eta(\theta^{[i]}_{k^{[i]}_{fs}})-\eta_*$ in part (i). Then in part (ii) we further derive the probabilistic upper bound of $\eta(\theta^{[i]}_{\infty})-\eta_*$ leveraging the result of Pareto improvement in [T5] of in Theorem \ref{thm: pareto improvement}. In part (iii), we  extend the upper bound to $\eta(\theta^{[j]}_{\infty})-\eta_*$ for all $j\in\mathcal{V}$ leveraging the result of Almost-consensus in [T4] of in Theorem \ref{thm: pareto improvement}. Finally, we characterize the probability of $\theta^{[i]}_{k^{[i]}_{fs}}\in\Psi(\theta_*)$.

\emph{Part (i): Probabilistic upper bound of $\eta(\theta^{[i]}_{k^{[i]}_{fs}})-\eta_*$.}
Suppose  there exists $i\in\mathcal{V}$ such that $\theta^{[i]}_{k^{[i]}_{fs}}\in \Psi(\theta_*)$ for some $\theta_*\in\Theta_*$.
The definition of $k^{[i]}_{fs}$ renders that $\|z^{[i]}_{k^{[i]}_{fs}}\|<q^{[i]}$. Combining this with Lemma \ref{lemma: gradient estimate} renders that
\begin{align}\label{ineq: P ||nabla eta||}
P\Big(\|\nabla\eta(\theta^{[i]}_{k^{[i]}_{fs}})\|\leqslant q^{[i]}+\epsilon_1\Big)\geqslant1-\frac{(\sigma^{[i]})^2}{\epsilon_1^2}.
\end{align}
Combining \eqref{ineq: lemma 1 cited} with Cauchy-Schwartz inequality implies 
\begin{align}\label{ineq: ||theta-theta*||<= ||nabla eta||}
	\alpha\|\theta-\theta_*\|\leqslant \| \nabla\eta(\theta)\|, ~\forall\theta\in\mathcal{K}(\theta_*).
\end{align}

Since $\theta^{[i]}_{k^{[i]}_{fs}}\in \Psi(\theta_*)\subset\mathcal{K}(\theta_*)$, combining \eqref{ineq: P ||nabla eta||} with inequality \eqref{ineq: ||theta-theta*||<= ||nabla eta||} renders
\begin{align*}
P\Big(\|\theta^{[i]}_{k^{[i]}_{fs}}-\theta_*\|\leqslant\frac{q^{[i]}+\epsilon_1}{\alpha}\mid\theta^{[i]}_{k^{[i]}_{fs}}\in \Psi(\theta_*)\Big)\geqslant
1-\frac{(\sigma^{[i]})^2}{\epsilon_1^2}.
\end{align*}
Combining this with Assumption \ref{assmp: lipschitz} further renders
\begin{align}\label{ineq: P eta theta fs - eta*}
&P\Big(\eta(\theta^{[i]}_{k^{[i]}_{fs}})-\eta_*\leqslant\frac{L_\eta(q^{[i]}+\epsilon_1)}{\alpha}\mid \theta^{[i]}_{k^{[i]}_{fs}}\in \Psi(\theta_*)\Big)\nonumber\\
&\geqslant1-\frac{(\sigma^{[i]})^2}{\epsilon_1^2}.
\end{align}

\emph{Part (ii): Probabilistic upper bound of $\eta(\theta^{[i]}_{\infty})-\eta_*$.}
Denote $\delta^{[i]}\triangleq\eta(\theta^{[i]}_\infty)-\eta(\theta^{[i]}_{k^{[i]}_{fs}})$. Notice that 
the definition of $J_E$ renders that $J_E\in[0,1]$. Then the definition of $\eta$ renders that $\eta\in[0,1]$.
Then it holds that $\delta^{[i]}\in[-1,1]$. Theorem \ref{thm: pareto improvement} [T3] implies that $\mathbb{E}[\delta^{[i]}\mid \theta^{[i]}_\infty\neq\theta^{[i]}_{k^{[i]}_{fs}}]\leqslant -2b^{\min}_\gamma$. Then let $\epsilon_2>0$, by leveraging Hoeffding's inequality in Theorem \ref{thm: hoeffding}, we have
\begin{align*}
&	P\Big(\eta(\theta^{[i]}_\infty)-\eta(\theta^{[i]}_{k^{[i]}_{fs}})\geqslant\epsilon_2\Big)\\
&\leqslant P\Big(\eta(\theta^{[i]}_\infty)-\eta(\theta^{[i]}_{k^{[i]}_{fs}})\geqslant\epsilon_2\mid \theta^{[i]}_\infty\neq\theta^{[i]}_{k^{[i]}_{fs}}\Big)\\
&\leqslant P\Big(\delta^{[i]}-\mathbb{E}[\delta^{[i]}| \theta^{[i]}_\infty\neq\theta^{[i]}_{k^{[i]}_{fs}}]\geqslant\epsilon_2+2b^{\min}_\gamma\mid \theta^{[i]}_\infty\neq\theta^{[i]}_{k^{[i]}_{fs}}\Big)\\
&\leqslant2\exp\big(-2(\epsilon_2+2b^{\min}_\gamma)^2\big)\leqslant2\exp\big(-2\epsilon_2^2\big).
\end{align*}
Combining this with \eqref{ineq: P eta theta fs - eta*} renders that
\begin{align}\label{prob: eta theta i infty - eta*}
	\eta(\theta^{[i]}_{\infty})-\eta_*\leqslant\frac{L_\eta(q^{[i]}+\epsilon_1)}{\alpha}+\epsilon_2
\end{align}
with probability at least 
$
(1-\frac{(\sigma^{[i]})^2}{\epsilon_1^2})(1-2\exp\big(-2\epsilon_2^2\big))\geqslant1-\frac{(\sigma^{[i]})^2}{\epsilon_1^2}-2\exp\big(-2\epsilon_2^2\big),
$ given $\theta^{[i]}_{k^{[i]}_{fs}}\in \Psi(\theta_*)$.

\emph{Part (iii): Probabilistic upper bound of $\eta(\theta^{[j]}_{\infty})-\eta_*$ for all $j\in\mathcal{V}$.}
Denote $\delta_\infty\triangleq \max_{j\in\mathcal{V}} \allowbreak \eta(\theta^{[j]}_\infty)-\min_{j\in\mathcal{V}}\eta(\theta^{[j]}_\infty)$. It is obvious that $\delta_\infty\geqslant 0$. Then combining Markov inequality with Theorem \ref{thm: pareto improvement} [T4], we have
\begin{align}
P\Big(\delta_\infty\geqslant2\epsilon_3b^{\max}_\gamma\Big)\leqslant\frac{1}{\epsilon_3}.
\end{align}

Combining this with \eqref{prob: eta theta i infty - eta*} renders that, given there exists $i\in\mathcal{V}$ such that $\theta^{[i]}_{k^{[i]}_{fs}}\in \Psi(\theta_*)$, it holds that, for all $j\in\mathcal{V}$,
\begin{align}\label{prob: eta theta infty - eta*}
	&\eta(\theta^{[j]}_{\infty})-\eta_*\leqslant\frac{L_\eta(q^{[i]}+\epsilon_1)}{\alpha}+\epsilon_2+2\epsilon_3b^{\max}_\gamma
\end{align}
with probability at least $1-\frac{(\sigma^{[i]})^2}{\epsilon_1^2}-2\exp(-2\epsilon_2^2)-\frac{1}{\epsilon_3}.$ 

\emph{Part (iv): Probability of there exists $i\in\mathcal{V}$ such that $\theta^{[i]}_{k^{[i]}_{fs}}\in\Psi(\theta_*)$.}
Given Assumption \ref{assmp: ||z||} holds, Theorem 4 in \cite{mertikopoulos2020almost} indicates that for each $\theta_*\in\Theta_*$, it holds that
\begin{align}\label{prob: thm4 cited}
	&P\Big(\theta^{[i]}_{k^{[i]}_{fs}}\in \Psi(\theta_*)\mid \theta^{[i]}_0\in\Psi_1(\theta_*)\Big)\geqslant1-\frac{R_*(\theta_*;\sigma^{[i]})\Gamma}{\epsilon_0(\theta_*)},
\end{align}
where $R(\theta_*;\sigma^{[i]})\triangleq L_{\eta}^2+(1+(4\epsilon_0(\theta_*)+2\sqrt{\epsilon_0(\theta_*)})^2)(\sigma^{[i]})^2$ and $\Gamma\triangleq r^{[i]}\sum_{k=1}^{\infty}\frac{1}{k^{2\rho}}$.

Denote $\omega\triangleq \frac{\beta(\Theta_0\cap[\cup_{\theta_*\in\Theta_*}\Psi_1(\theta_*)])}{\beta(\Theta_0)}$. Since $\beta(\Theta_0\cap[\cup_{\theta_*\in\Theta_*}\Psi_1(\theta_*)])>0$, it is obvious that $\omega\in(0,1]$.
Since $\theta^{[i]}_0$ is uniformly sampled over compact set $\Theta_0$, we have $P\Big(\theta_0^{[i]}\in\Psi_1(\theta_*)\mid \theta_*\in\Theta_*\Big)=\omega$. Since there are $|\mathcal{V}|$ learners in $\mathcal{V}$ and $\theta^{[i]}_0$ are independently sampled for all $i\in\mathcal{V}$, then we further have
\begin{align}\label{prob: in neighborhood}
	&P\Big(\exists i\in\mathcal{V}\text{ such that }\theta_0^{[i]}\in\Psi_1(\theta_*)\mid \theta_*\in\Theta_*\cap\Theta_0\Big)\nonumber\\
	&=1-P\Big(\theta_0^{[i]}\not\in\Psi_1(\theta_*;\epsilon),~\forall i\in\mathcal{V}\mid \theta_*\in\Theta_*\cap\Theta_0\Big)\nonumber\\
	&=1-(1-\omega)^{|\mathcal{V}|}.
\end{align}
Combining \eqref{prob: thm4 cited} with \eqref{prob: in neighborhood} renders 
\begin{align}\label{prob: exists in neighborhood}
	&P\Big(\exists i\in\mathcal{V}\text{ such that }\theta_\infty^{[i]}\in\Psi(\theta_*)\mid \theta_*\in\Theta_*\cap\Theta_0\Big)\nonumber\\
&\geqslant1-(1-\omega)^{|\mathcal{V}|}-\max_{\theta_*\in\Theta_*}\frac{R_*(\theta_*;\sigma^{\max})\Gamma}{\epsilon_0(\theta_*)}.
\end{align}

Combining \eqref{prob: exists in neighborhood} with \eqref{prob: eta theta infty - eta*} concludes the proof. $\hfill\blacksquare$

\section{Simulation}\label{sec: simulation}
In this section, we conduct a set of Monte Carlo simulations to evaluate the performance of the FedGen algorithm in the PyBullet simulator \cite{pybullet}. All the simulations are conduct in Python on an Intel Core i5 CPU, 4.10 GHz, with 16 GB of RAM.

\begin{figure}
	\centering
	\includegraphics[width=0.4\textwidth]{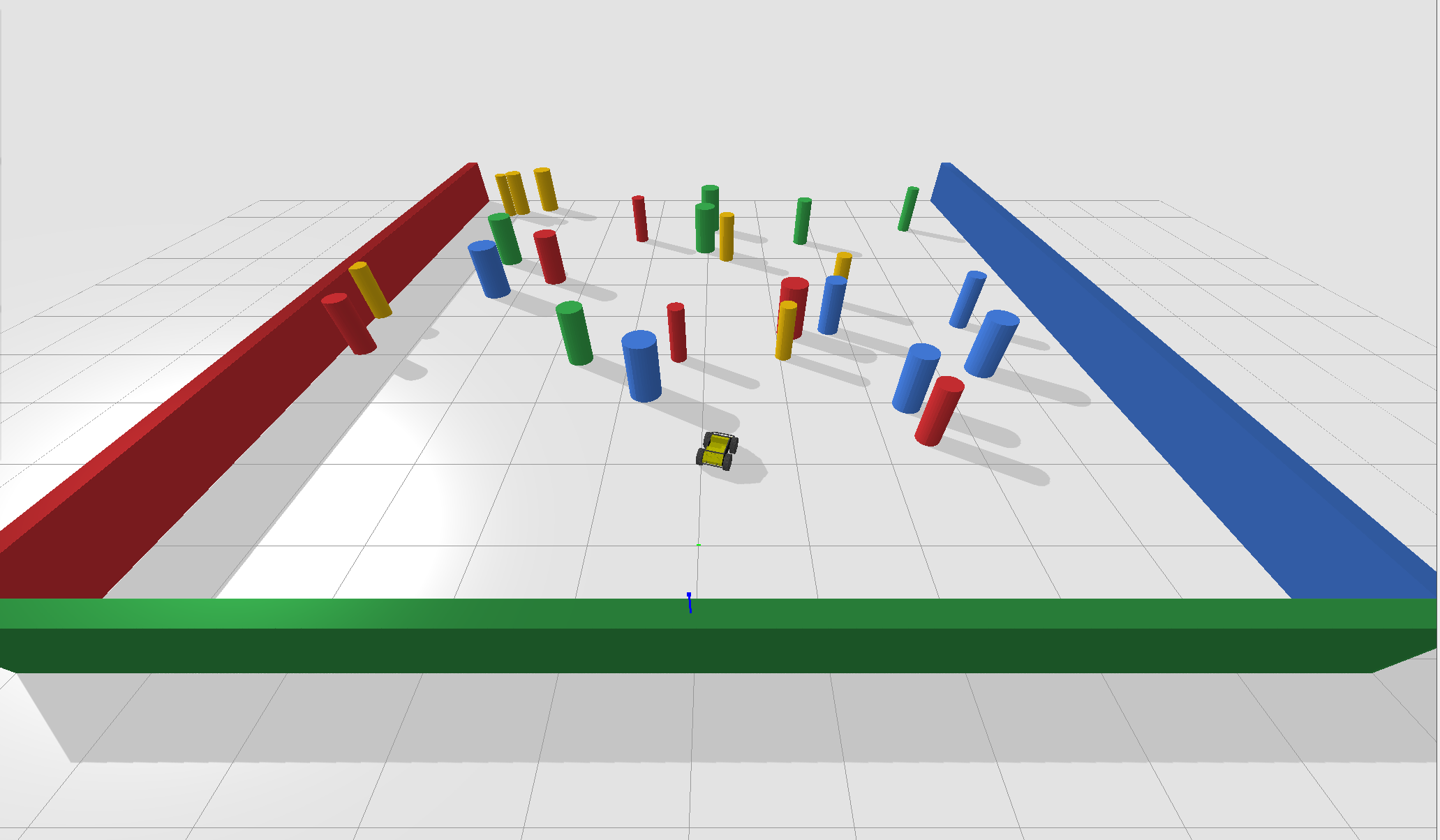}
	\caption{A sample environment in PyBullet}
	\label{fig:env}
\end{figure}

{\em (Environment configuration)}.
The evaluation is conducted using Zermelo's navigation problem \cite{Zlobec2001} in a 2D space, where the environments are randomly generated.  A sample of the environments is shown in Figure \ref{fig:env}.
Each environment $E$ consists of $n_{obs}$ cylinder obstacles and
three walls as the boundary of the 2D environment 
with horizontal coordinate $x_1 \in [-5,5]$ and vertical coordinate $x_2 \in [0,10]$.
The environments are generated by sampling the obstacle number $n_{obs}$ uniformly between $15$ and $30$, and then independently sampling the centers of the cylinders from a uniform distribution over the ranges $[-5,5]\times[2,10]$. The radius of each obstacle is sampled independently from a uniform distribution over  $[0.1,0.25]$. 
The goal of the robot is to reach the open end of the environment while avoiding collision with the walls and the obstacles.  

{\em (Robot dynamics).}
We consider a four-wheel robot with  constant speed $v=2.5$ and length $L=0.08$ subject to unknown environment-specific  disturbances $d_E$. The dynamics of the robot with state $x=[x_1,x_2,x_3]$ is given by
$
\dot{x}_1=v\cos(x_3)+d_E(x_1,x_2)$, $\dot{x}_2=v\sin(x_3)$, $\dot{x_3}=\tan(u)/L$,
where $x_3$ is the heading of the robot, control $u \in [-0.25\pi, 0.25\pi]$, and $d_E$ is generated using the Von Karman power spectral density function as described in \cite{cole2013impact} representing the  road texture disturbance (e.g., bumps and slippery surface) in environment $E$. 



{\em  (Sensor model). } In the simulation, the robots are equipped with a sensor able to obtain the robot's state information $x$ and a depth sensor (e.g., LiDAR) able to measure the distances between the robot and the obstacles. The sensors are perfect.
 The readings of the depth sensor depend on the environment $E$ and the state of the robot. Specifically, the output of the sensor has $20$ entries, where each entry $\phi$ corresponds to the distance measurement at angle $x_3-\pi / 3+ (\phi-1)\pi / 60$ with $\phi=1,\cdots, 20$.  The  measurement $h_{\phi}(x,\mathcal{X}_{O,E})$ provides the shortest distance  between the  obstacles, if there is any, at the angle of entry $\phi$ of the robot    and the robot at location $(x_1,x_2)$. The sensing range is $5$, i.e., $h_{\phi}(x,\mathcal{X}_{O,E})\in[0,5]$. That is, the observation function is given by $h(x,\mathcal{X}_{O,E})=[x,h_{1}(x,\mathcal{X}_{O,E}),\cdots, h_{20}(x,\mathcal{X}_{O,E})]$.  

\subsection{Training}

We consider a deep neural network-based control policy
$
\pi_\theta,
$
that is parameterized by $\theta$, the weights of the neural network.
Note that the control policy is periodic in $\varphi$. Thus, the input $\varphi$
is replaced by two inputs $\sin(\varphi)$ and $\cos(\varphi)$. 
During training, especially during the early phase, the original cost functional $ J_E(x_{int},\pi_\theta)$ may have zero gradient for some initial state $x_{int}$ since collisions with obstacles dominate most of the  trial runs.
Therefore, to facilitate training, we consider the surrogate
$\hat{J}_E(x_{int},\theta)\triangleq 0.1 \rho_E(x_{int},\pi_\theta)+J_E(x_{int},\pi_\theta)$, where  $\rho_E(x_{int},\pi_\theta)\triangleq\min_{x_G\in\mathcal{X}_{G,E}}\|x(t_{end}(x_{int},\pi_\theta;E))-x_G\|$ is the distance between the location of the first collision and the goal region.
The cost $\rho_E(x_{int},\pi_\theta)$ is to drive the robot approaching the goal without collision, and the cost $J_E(x_{int},\pi_\theta)$ is to minimize the arrival time when the robot is able to safely reach the goal.

Since it is challenging to derive the analytical expression of $\nabla \hat{J}_E(x_{int},\theta)$, we approximate it by natural evolution strategies \cite{wierstra2014natural,sehnke2010parameter}. 
In particular, 
we suppose $\theta$ follows
a multivariate Gaussian distribution such that $\theta\sim\mathcal{N}(\mu, \Sigma)$ with mean $\mu \in \mathbb{R}^{n_\theta}$ and diagonal covariance $\Sigma \in \mathbb{R}^{n_\theta \times n_\theta}$. Let $\sigma \in \mathbb{R}^{n_\theta}$ be a vector aggregating the square-root of the diagonal elements of $\Sigma$.
The gradients of ${\mathbb{E}}_{\theta}\left[\hat{J}_E(x_{int},\pi_\theta)\right]$ with respect to $\mu$ and $\sigma$ are
$$
\begin{aligned}
	&\nabla_{\mu} \underset{\theta \sim \mathcal{N}(\mu,\Sigma)}{\mathbb{E}}\left[\hat{J}_E(x_{int},\pi_\theta)\right]=\\
	&\quad\underset{\epsilon \sim \mathcal{N}(0, I)}{\mathbb{E}}\left[\hat{J}_E(x_{int},\pi_{\mu+\sigma \odot \epsilon}) \epsilon\right] \oslash \sigma, \\
	&\nabla_{\sigma} \underset{\theta \sim \mathcal{N}(\mu,\Sigma)}{\mathbb{E}}\left[\hat{J}_E(x_{int},\pi_\theta)\right]=\\
	&\quad\underset{\epsilon \sim \mathcal{N}(0, I)}{\mathbb{E}}\left[\hat{J}_E(x_{int},\pi_{\mu+\sigma \odot \epsilon})(\epsilon \odot \epsilon-\mathbf{1})\right] \oslash \sigma,
\end{aligned}
$$
where $\oslash$ is the element-wise division, $\odot$ is the elementwise product, and $\mathbf{1}$ is a vector of 1's with dimension $n_\theta$. 
We approximate the expectation by  collecting 30 samples of ${\epsilon \sim \mathcal{N}(0, I)}$ and taking the average.
To reduce the variance in the expectation approximation, antithetic sampling  \cite{salimans2017evolution} is employed. That is, the update of $\theta$ is then replaced by the updates of $\mu$ and $\sigma$, and $\mu$ is returned as the estimate of $\theta$.

{\em (Selection of hyperparameters).}
The neural network control policy consists of an input layer
of size 24, followed by 3 hidden layers of size 20 with ReLu
nonlinearities and an output layer of size 1.
We pick $n^{[i]}_\mathcal{E}=10$, $n^{[i]}_{int|\mathcal{E}}=1$, $\gamma=0.01$, $r=0.01$, $L_\eta=0.1$,  $q^{[i]}=0.04$, and $8$ learners, i.e., $|\mathcal{V}|=8$, for the experiments. 
The generalized performance in unseen environments is defined as an expectation over all possible environments, which cannot be obtained exactly. Therefore, we estimate the  generalized performances using $10^4$ sample environments.

\subsection{Results}

\begin{figure}[thpb]
	\centering
	\begin{subfigure}[t]{0.235\textwidth}
		\centering
		\includegraphics[width=1.0\textwidth]{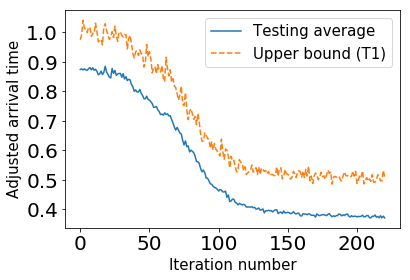}
		\caption{Expected normalized arrival time}
	\end{subfigure}
	\hfil
	\begin{subfigure}[t]{0.235\textwidth}
		\centering
		\includegraphics[width=1.0\textwidth]{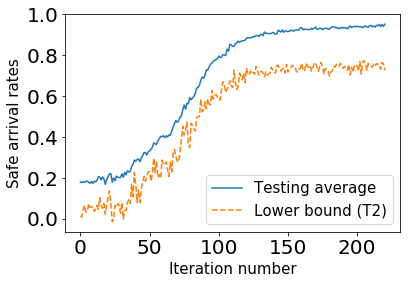}
		\caption{Safe arrival rate}
	\end{subfigure}
	\caption{Generalized performances to unseen environments}
	\label{fig: monotonic improvement}
\end{figure}

{\em (Generalization and convergence)}. Figure \ref{fig: monotonic improvement} compares the upper bound on the expected normalized arrival time (T1) and the lower bound on the safe arrival rate (T2) in Theorem \ref{thm: monotonic optimality} respectively with the actual expected normalized arrival time  and the actual safe arrival rate  of learner $1$. Other learners have similar behaviors. As the figure illustrates, the upper bound and the lower bound derived in the theorem are valid. This shows that the control policy trained  can zero-shot generalize well to the $10^4$  unseen environments. Converging behavior is also obvious in Figure \ref{fig: monotonic improvement}, which aligns with (T3) of Theorem \ref{thm: pareto improvement}.

{\em (Near consensus and Pareto improvement).} In Table \ref{tab:Learner-wise near optimality} below, we show the performances of the learners' estimates in terms of  
the  expected distance-to-goal  $0.1\rho_E$, the expected  normalized arrival time $J_E$, and the expected safe arrival rate. We compare with the control policy at  initialization ($\theta^{[i]}_0$), the control policy obtained without communication ($\theta^{[i]}_{k^{[i]}_{fs}}$), i.e., the control policy obtained by running FedGen using $\mathcal{V}=\{i\}$, and the final convergence ($\theta^{[i]}_\infty$) under FedGen.  We can observe that all the expected costs, expected normalized arrival times and expected safe arrival rates at $\theta^{[i]}_\infty$ are roughly equal. This aligns with the almost consensus (T4) in Theorem \ref{thm: pareto improvement}. Furthermore, we can observe that all the expected costs and the expected normalized arrival times at  $\theta^{[i]}_\infty$ are no larger than those of $\theta^{[i]}_0$ and $\theta^{[i]}_{k^{[i]}_{fs}}$, while the expected safe arrival rates at $\theta^{[i]}_\infty$ are no smaller than those at $\theta^{[i]}_0$ and $\theta^{[i]}_{k^{[i]}_{fs}}$. This shows that FedGen  brings Pareto improvement for each learner through communication, which is also shown in (T5) of Theorem \ref{thm: pareto improvement}.

{\em (Performance vs. the number of learners).}
Table \ref{tab: performance vs learner number} presents the expected distance-to-goal, normalized arrival time, and safe arrival rate of the limiting estimate $\theta^{[i]}_\infty$ when FedGen is run using different number of learners. The table shows that with more learners involved in FedGen, the performances of the control policies are better. This shows a stronger result than that in Theorem \ref{thm: global convergence}, where more learners can only improve the probability of achieving the optimality gap in \eqref{ineq: optimality gap}.

\begin{figure*}[thpb]
	\centering
	\begin{subfigure}[t]{0.3\textwidth}
		\centering
		\includegraphics[width=1\textwidth]{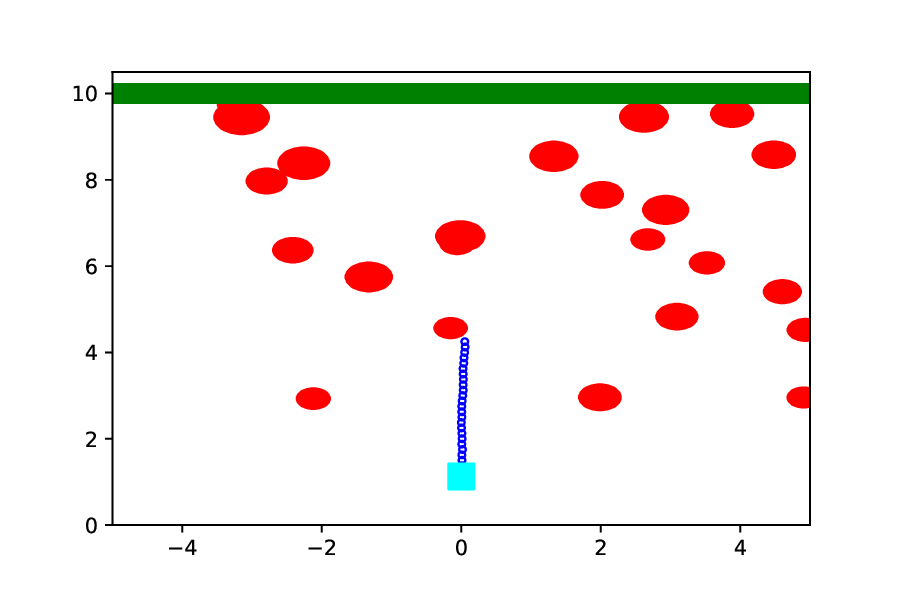} 
		\caption{Trajectory produced by $\theta^{[1]}_0$}
		\label{fig:initial policy}
	\end{subfigure}
	\hfil
	\begin{subfigure}[t]{0.3\textwidth}
		\centering
		\includegraphics[width=1\textwidth]{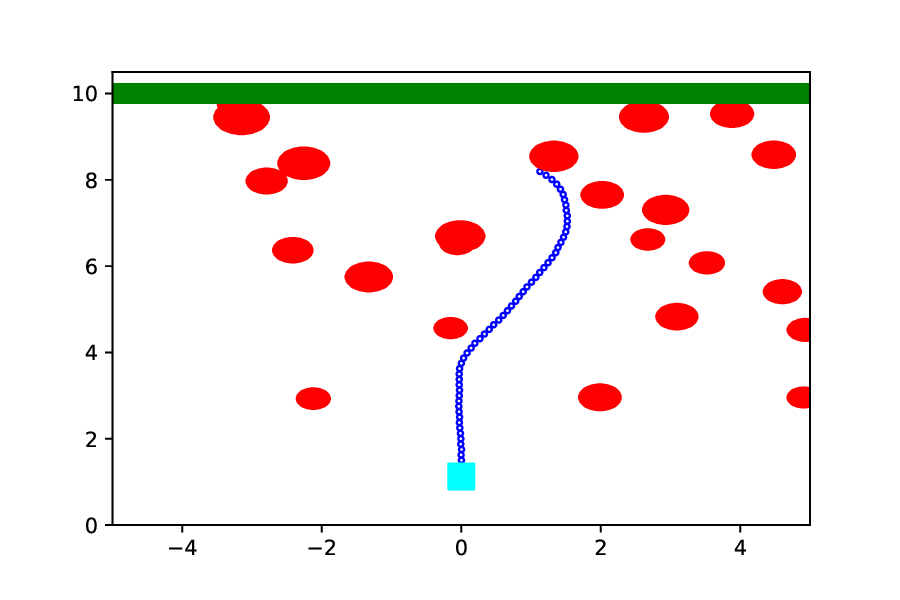}
		\caption{Trajectory produced by $\theta^{[1]}_{k^{[1]}_{fs}}$}
		\label{fig:local policy}
	\end{subfigure}
	\hfil
	\begin{subfigure}[t]{0.3\textwidth}
		\centering
		\includegraphics[width=1\textwidth]{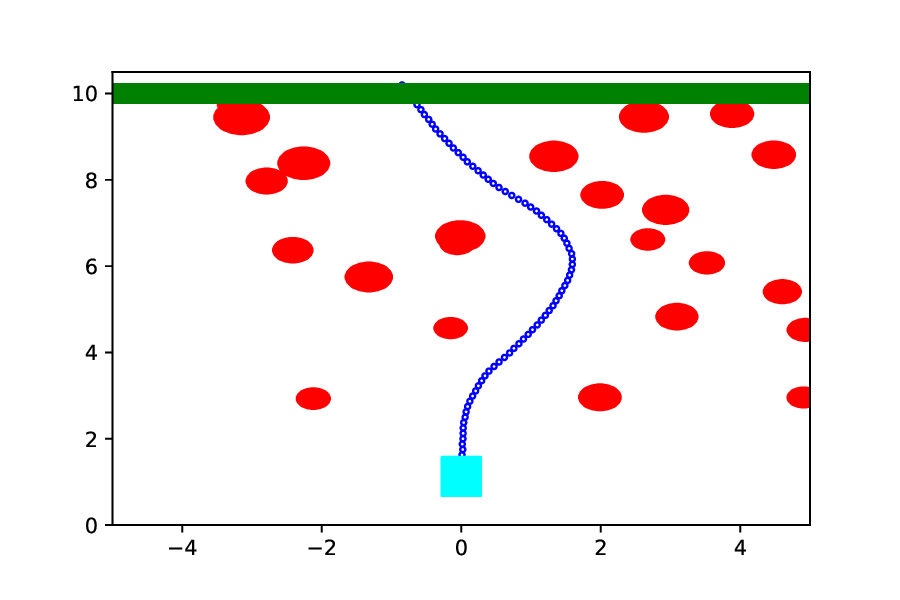}
		\caption{Trajectory produced by $\theta^{[1]}_\infty$}
		\label{fig: final policy}
	\end{subfigure}
	\caption{Comparison between initial policy, locally converged policy and globally converged policy}
	\label{fig: compare policy}
\end{figure*}

Graphically,  Figure \ref{fig: compare policy} respectively shows the trajectories of  the robot in a sample of unseen environments using learner $1$'s initial policy $\theta^{[1]}_0$, locally converged policy $\theta^{[1]}_{k^{[1]}_{fs}}$ and finally converged policy $\theta^{[1]}_\infty$. The red disks represent the obstacles. The cyan square represents the initial location. The green line represents the goal region. The blue curves are the trajectories of the robot.
Both the initial control policy (Figure \ref{fig:initial policy}) and the locally converged control policy (Figure \ref{fig:local policy}) cannot bring the robot to the open end, despite the locally converged control policy is able to drive the robot closer to the open end.
Nevertheless, the path generated by the final control policy $\theta^{[1]}_\infty$ is able to drive the robot to the open end. This illustrates that FedGen helps the learners escape from their local minima and  achieve better generalizability.
Additional figures with other realizations of the environments can be found in Appendix \ref{appdix}.
\begin{table*}[t]
	\centering
	\begin{tabular}{|P{3.2cm}|c|c|c|c|c|c|c|c|c|}
		\hline
		\multicolumn{2}{|c|}{Learner ID ($i$)}   & 1&2&3&4&5&6&7&8 \\
		\hline
		\multirow{3}{*}{\thead{Distance-to-goal\\ $\big(0.1\mathbb{E}[\rho_E(x_{int},\pi_{\theta^{[i]}})]\big)$}}   &Init($\theta^{[i]}_0$)& 0.5198 &0.5170&0.5208&0.5210&0.5148&0.5231&0.5237&0.5167 \\
		&Local($\theta^{[i]}_{k^{[i]}_{fs}}$)&0.0436&{\bf 0.0396}&{\bf 0.0331}&0.4810&0.4105&{\bf 0.0341}&0.3992&0.4989\\
		&{\bf Final($\theta^{[i]}_\infty$)}&{\bf 0.0374}&{\bf 0.0396}&{\bf 0.0331}&{\bf 0.0335}& {\bf 0.0353} &{\bf 0.0341}&{\bf 0.0363}&{\bf 0.0335}\\
		\hline
		\multirow{3}{*}{\thead{Normalized arrival time\\ $\big(\mathbb{E}[J_E(x_{int};\pi_{\theta^{[i]}})]\big)$}} &Init($\theta^{[i]}_0$)&0.8743&0.8761& 0.8744&0.8782&0.8692&0.8748&0.8797&0.8732\\
		&Local($\theta^{[i]}_{k^{[i]}_{fs}}$)&0.3759&{\bf 0.3763}&{\bf 0.3701}&0.8385&0.7815&{\bf 0.3700}&0.7622&0.8569\\
		&{\bf Final($\theta^{[i]}_\infty$)}&{\bf 0.3748}&{\bf 0.3763}&{\bf 0.3701}&{\bf 0.3679}&{\bf 0.3711}&{\bf 0.3700}&{\bf 0.3716}&{\bf 0.3704}\\
		\hline
		\multirow{3}{*}{Safe arrival rate} &Init($\theta^{[i]}_0$)&0.1802&0.1776&0.1800&0.1746&0.1876&0.1794&0.1724&0.1818\\
		&Local($\theta^{[i]}_{k^{[i]}_{fs}}$)&0.9320&{\bf 0.9408}&{\bf 0.9522}&0.2314&0.3172&{\bf 0.9450}&0.3426&0.2054\\
		&{\bf Final($\theta^{[i]}_\infty$)}&{\bf 0.9386}&{\bf 0.9408}&{\bf 0.9522}&{\bf 0.9452}&{\bf 0.9432}&{\bf 0.9450}&{\bf 0.9428}&{\bf 0.9468}\\
		\hline
	\end{tabular}
	\caption{The  expected distance-to-goal, normalized arrival times, safe arrival rates of the estimates at initialization, local convergence and final convergence.}
	\label{tab:Learner-wise near optimality}
\end{table*}

\begin{table*}[t]
	\centering
	\begin{tabular}{|c|c|c|c|c|c|}
		\hline
	Number of learners ($|\mathcal{V}|$)  & 1&2&4&6&8 \\
		\hline
\thead{Distance-to-goal\\ $\big(0.1\mathbb{E}[\rho_E(x_{int},\pi_{\theta^{[i]}_{\infty}})]\big)$}&0.4989&$0.1548\pm0.0132$&$0.1391\pm 0.0247$&$0.0760\pm0.0126$&$0.0354\pm0.0021$\\
		\hline
\thead{	Normalized arrival time\\ $\big(\mathbb{E}[J_E(x_{int};\pi_{\theta^{[i]}_\infty})]\big)$} &0.8569&$0.4997\pm0.0158$&$0.4910\pm 0.0234$&$0.4111\pm 0.0169$&$0.3715\pm0.0027$\\
		\hline
		Safe arrival rate  &0.2054&$0.7325\pm0.0225$&$0.7563\pm 0.0338$&$0.8717\pm 0.0256$&$0.9442\pm0.0038$\\
		\hline
	\end{tabular}
	\caption{The  expected distance-to-goal, normalized arrival times, safe arrival rates of the limiting estimates for different number of learners. The table shows the average values over the learners plus-minus one standard deviation.}
	\label{tab: performance vs learner number}
\end{table*}

\section{Conclusion}
We propose FedGen, a federated reinforcement learning algorithm which allows a group of learners to collaboratively learn a single control policy for robot motion planning  with zero-shot generalization. 
The problem is formulated as an expected cost minimization problem and solved in a federated manner.
The proposed algorithm is able to provide zero-shot generalization guarantees on the performances of the local control policies in unseen environments as well as almost-sure convergence, almost consensus and Pareto improvement. The algorithm is evaluated using Monte Carlo simulations. Interesting future works include extensions to different objective functions and time-varying environments.

\bibliographystyle{abbrv}
\bibliography{Biblio-dataset-FedGen}

\begin{wrapfigure}{l}{0.11\textwidth}
	\centering
\includegraphics[width=1in,clip,keepaspectratio]{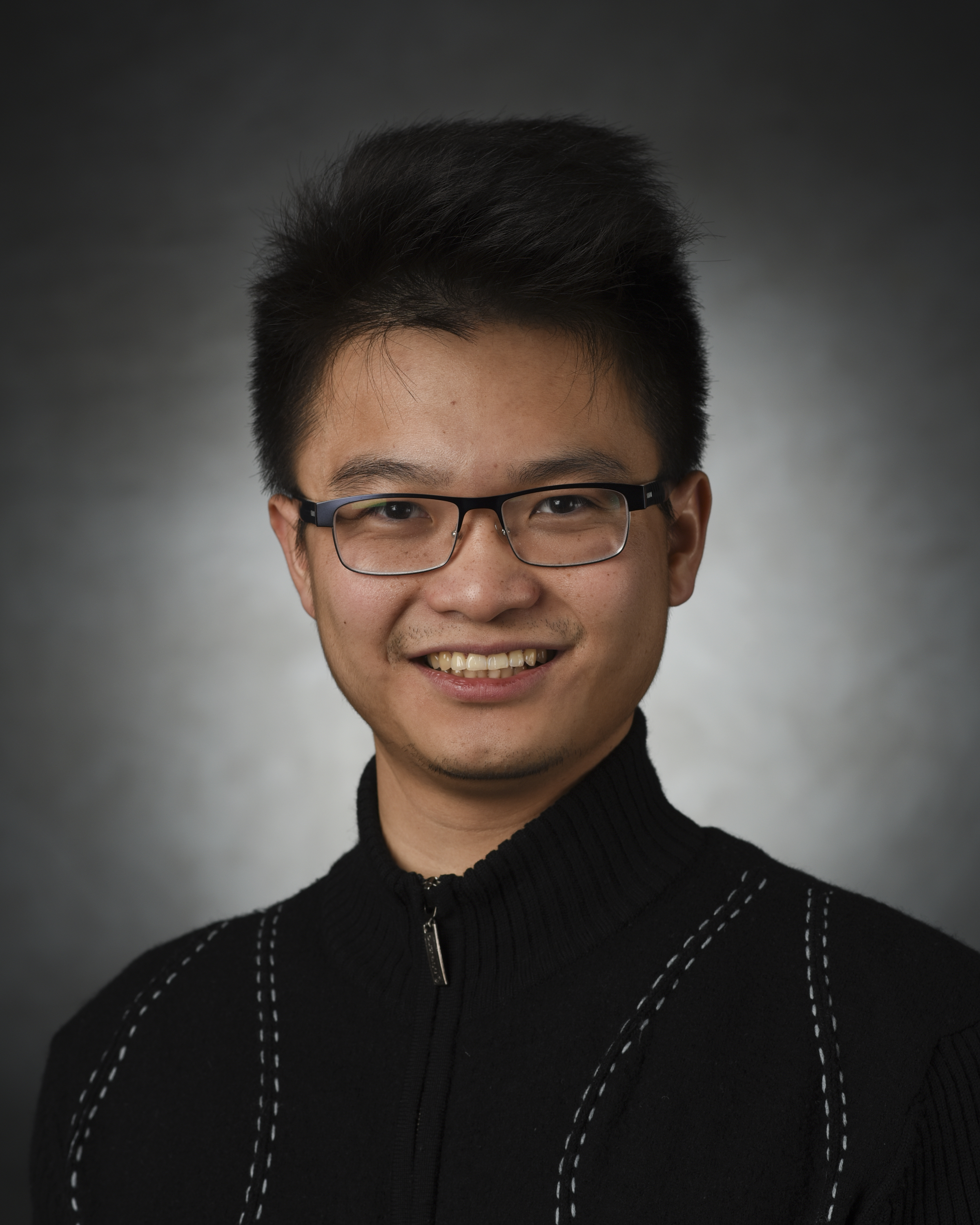}
\end{wrapfigure}\noindent
{\bf Zhenyuan Yuan} is  a  Ph.D.  candidate  in  the  School of  Electrical  Engineering  and  Computer  Science  at the  Pennsylvania  State  University.  He  received   B.S. in Electrical Engineering and B.S. in Mathematics from the  Pennsylvania  State  University  in  2018. His research interests lie in machine learning and motion planning with applications in robotic networks. He is a recipient of the Rudolf Kalman Best Paper Award of the ASME Journal of Dynamic Systems Measurement and Control in 2019 and the Penn State Alumni Association Scholarship for Penn State Alumni in the Graduate School in 2021.     


\begin{wrapfigure}{l}{0.11\textwidth}
	\centering
	\includegraphics[width=1in]{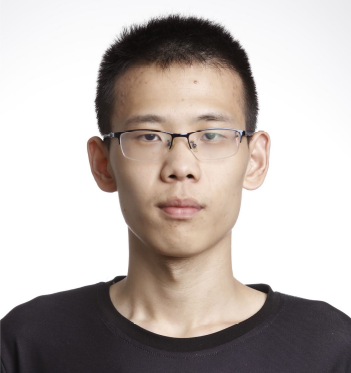}
\end{wrapfigure}\noindent
{\bf Siyuan Xu} is a Ph.D. candidate in the School of Electrical Engineering and Computer Science at the Pennsylvania State University. He received B.S. in Electrical Engineering from the Xi'an Jiaotong University in 2019. His research interests mainly focus on machine learning and motion planning.\par
\vspace{5em}
\begin{wrapfigure}{l}{0.11\textwidth}
	\centering
	\includegraphics[width=1in,clip,keepaspectratio]{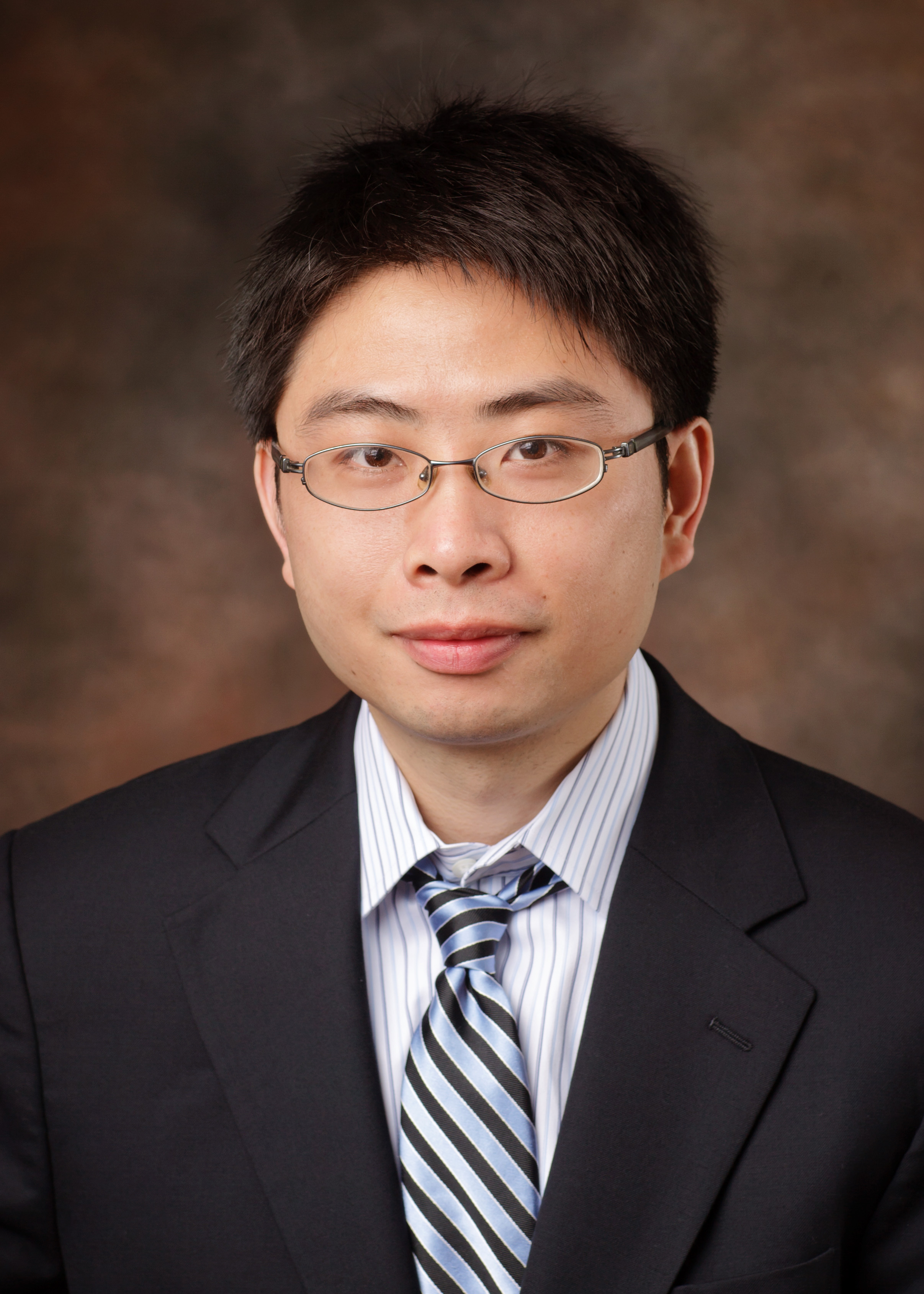}
\end{wrapfigure}\noindent
{\bf Minghui Zhu} is an Associate Professor in the School of Electrical Engineering and Computer Science at the Pennsylvania State University. Prior to joining Penn State in 2013, he was a postdoctoral associate in the Laboratory for Information and Decision Systems at the Massachusetts Institute of Technology. He received Ph.D. in Engineering Science (Mechanical Engineering) from the University of California, San Diego in 2011. His research interests lie in distributed control and decision-making of multi-agent networks with applications in robotic networks, security and the smart grid. He is the co-author of the book "Distributed optimization-based control of multi-agent networks in complex environments" (Springer, 2015). He is a recipient of the Dorothy Quiggle Career Development Professorship in Engineering at Penn State in 2013, the award of Outstanding Reviewer of Automatica in 2013 and 2014, and the National Science Foundation CAREER award in 2019. He is an associate editor of the IEEE Open Journal of Control Systems, the IET Cyber-systems and Robotics and the Conference Editorial Board of the IEEE Control Systems Society.

 \appendix
 \section{Performances of control policies  in other environments}\label{appdix} 
\begin{figure*}
 	\centering
 	\begin{subfigure}[t]{0.3\textwidth}
 		\centering
 		\includegraphics[width=1\textwidth]{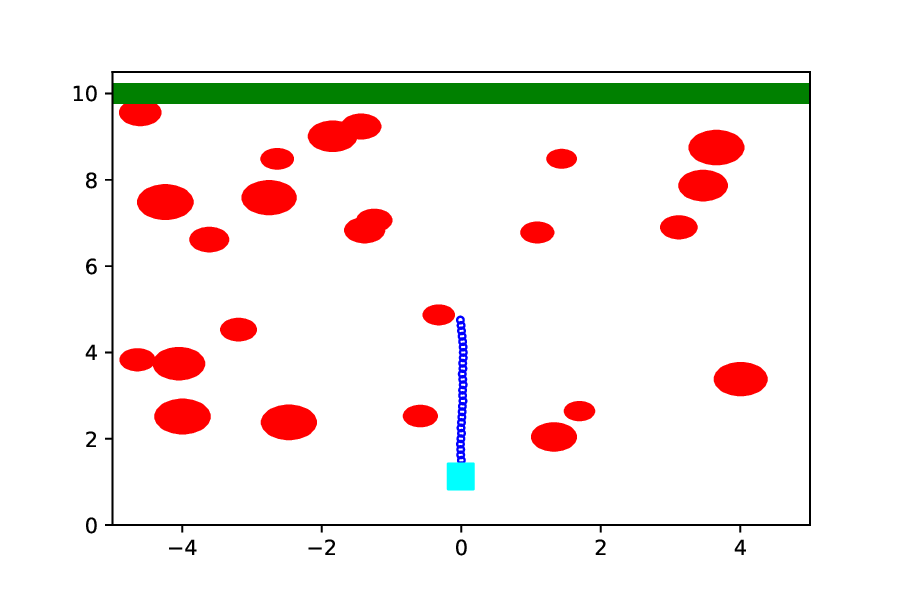} 
 		\caption{Trajectory produced by $\theta^{[1]}_0$}
 	\end{subfigure}
 	\hfil
 	\begin{subfigure}[t]{0.3\textwidth}
 		\centering
 		\includegraphics[width=1\textwidth]{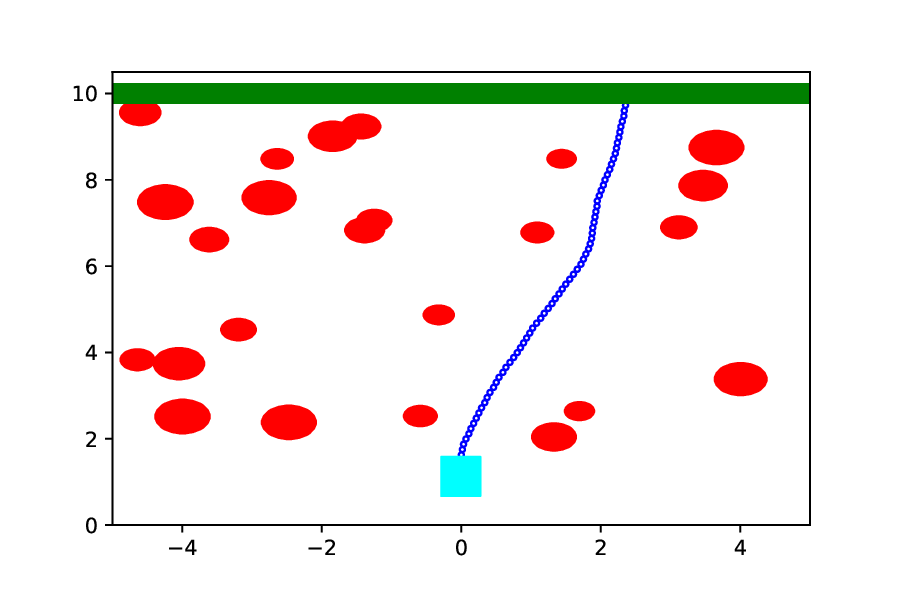}
 		\caption{Trajectory produced by $\theta^{[1]}_{k^{[1]}_{fs}}$}
 	\end{subfigure}
 	\hfil
 	\begin{subfigure}[t]{0.3\textwidth}
 		\centering
 		\includegraphics[width=1\textwidth]{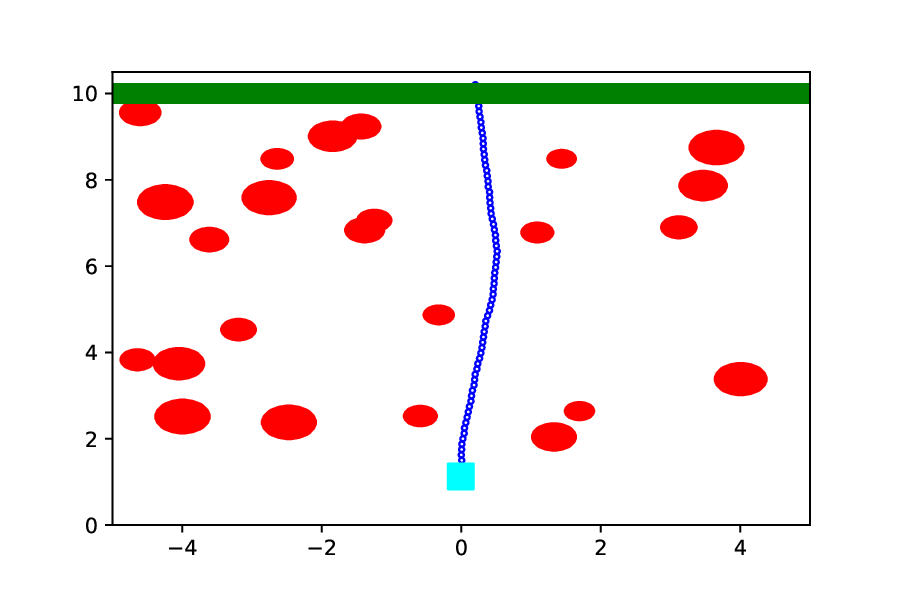}
 		\caption{Trajectory produced by $\theta^{[1]}_\infty$}
 	\end{subfigure}
 	\caption{Comparison between policies in  environment realization 2}
 	\label{fig: compare policy 2}
 \end{figure*}

\begin{figure*}
 	\centering
 	\begin{subfigure}[t]{0.3\textwidth}
 		\centering
 		\includegraphics[width=1\textwidth]{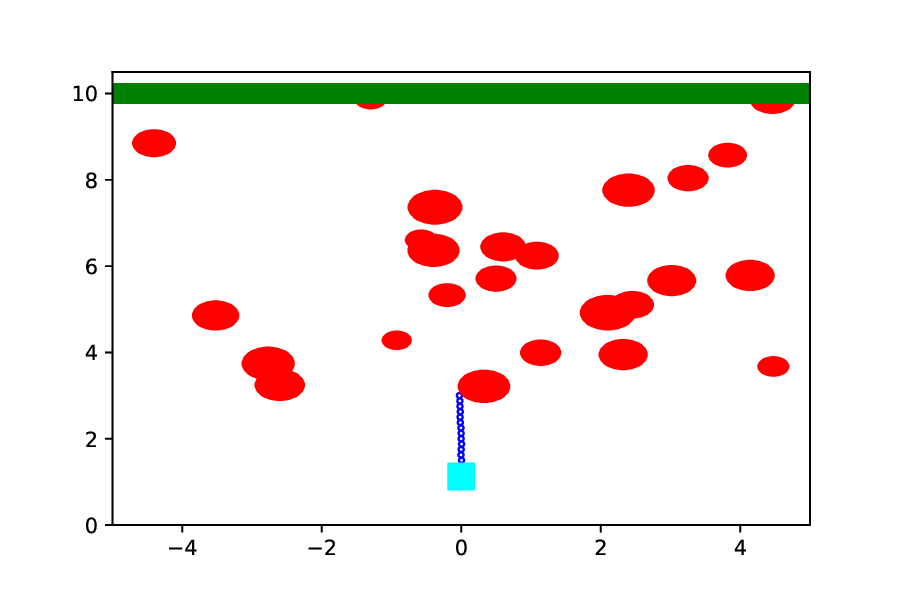} 
 		\caption{Trajectory produced by $\theta^{[1]}_0$}
 	\end{subfigure}
 	\hfil
 	\begin{subfigure}[t]{0.3\textwidth}
 		\centering
 		\includegraphics[width=1\textwidth]{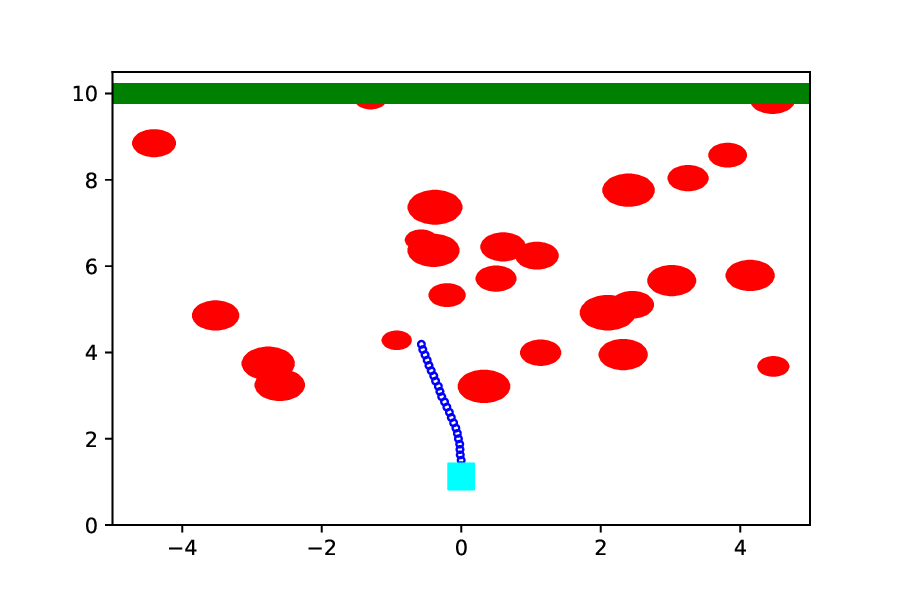}
 		\caption{Trajectory produced by $\theta^{[1]}_{k^{[1]}_{fs}}$}
 	\end{subfigure}
 	\hfil
 	\begin{subfigure}[t]{0.3\textwidth}
 		\centering
 		\includegraphics[width=1\textwidth]{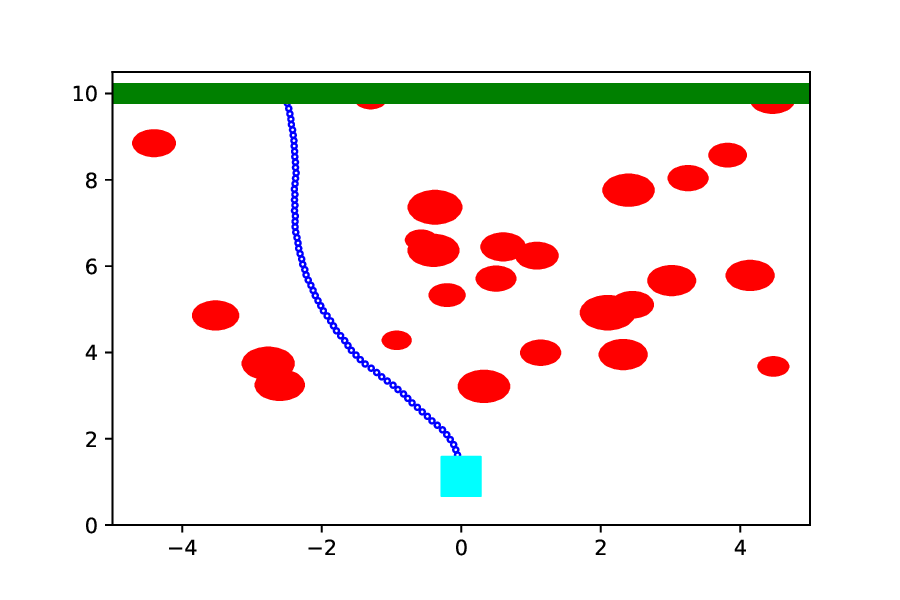}
 		\caption{Trajectory produced by $\theta^{[1]}_\infty$}
 	\end{subfigure}
 	\caption{Comparison between policies in  environment realization 3}
 	\label{fig: compare policy 3}
 \end{figure*}

\begin{figure*}
 	\centering
 	\begin{subfigure}[t]{0.3\textwidth}
 		\centering
 		\includegraphics[width=1\textwidth]{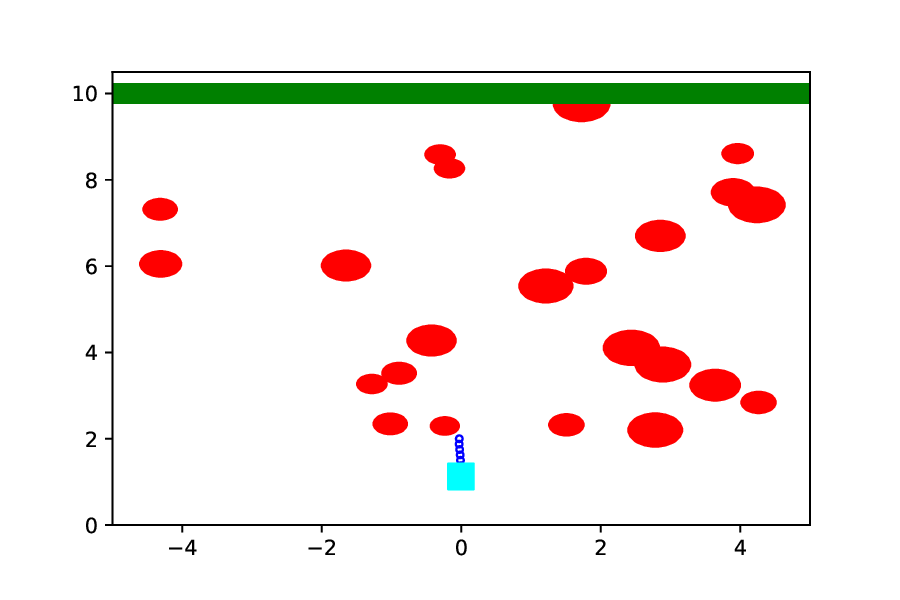} 
 		\caption{Trajectory produced by $\theta^{[1]}_0$}
 	\end{subfigure}
 	\hfil
 	\begin{subfigure}[t]{0.3\textwidth}
 		\centering
 		\includegraphics[width=1\textwidth]{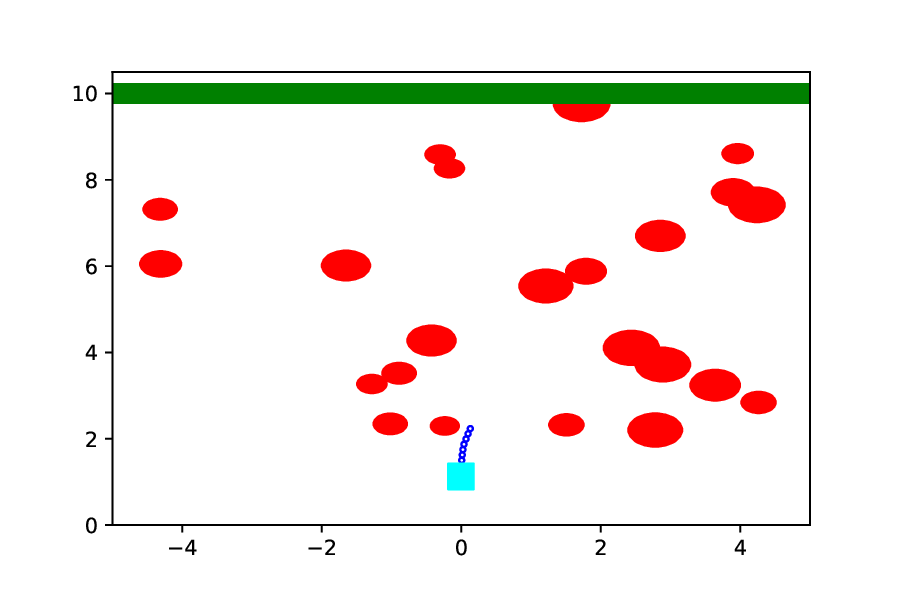}
 		\caption{Trajectory produced by $\theta^{[1]}_{k^{[1]}_{fs}}$}
 	\end{subfigure}
 	\hfil
 	\begin{subfigure}[t]{0.3\textwidth}
 		\centering
 		\includegraphics[width=1\textwidth]{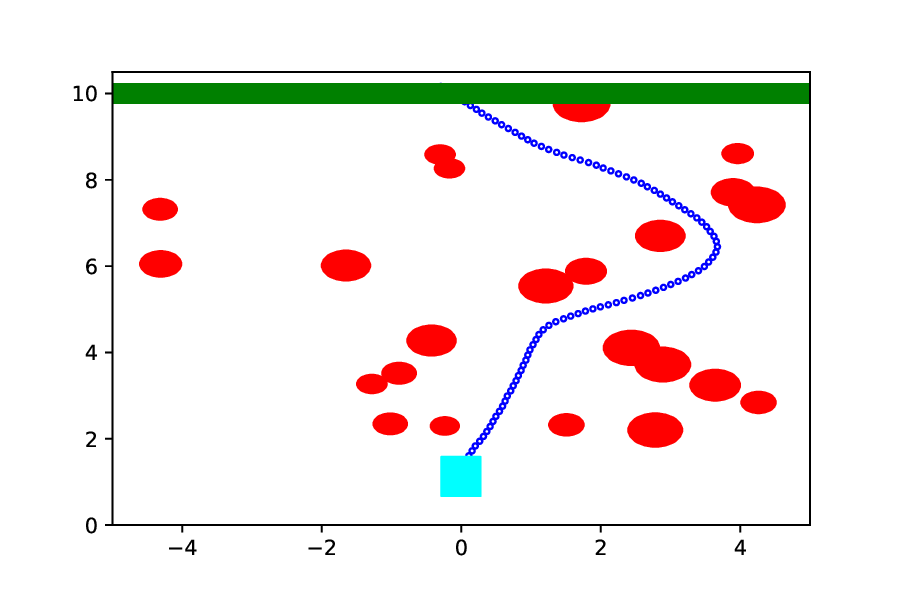}
 		\caption{Trajectory produced by $\theta^{[1]}_\infty$}
 	\end{subfigure}
 	\caption{Comparison between policies in  environment realization 4}
 	\label{fig: compare policy 4}
 \end{figure*}

\begin{figure*}
 	\centering
 	\begin{subfigure}[t]{0.3\textwidth}
 		\centering
 		\includegraphics[width=1\textwidth]{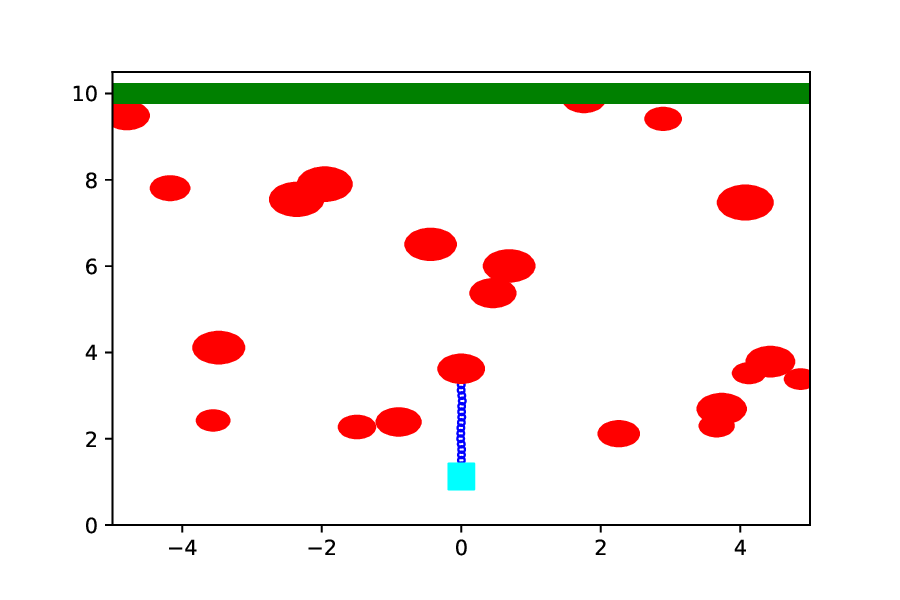} 
 		\caption{Trajectory produced by $\theta^{[1]}_0$}
 	\end{subfigure}
 	\hfil
 	\begin{subfigure}[t]{0.3\textwidth}
 		\centering
 		\includegraphics[width=1\textwidth]{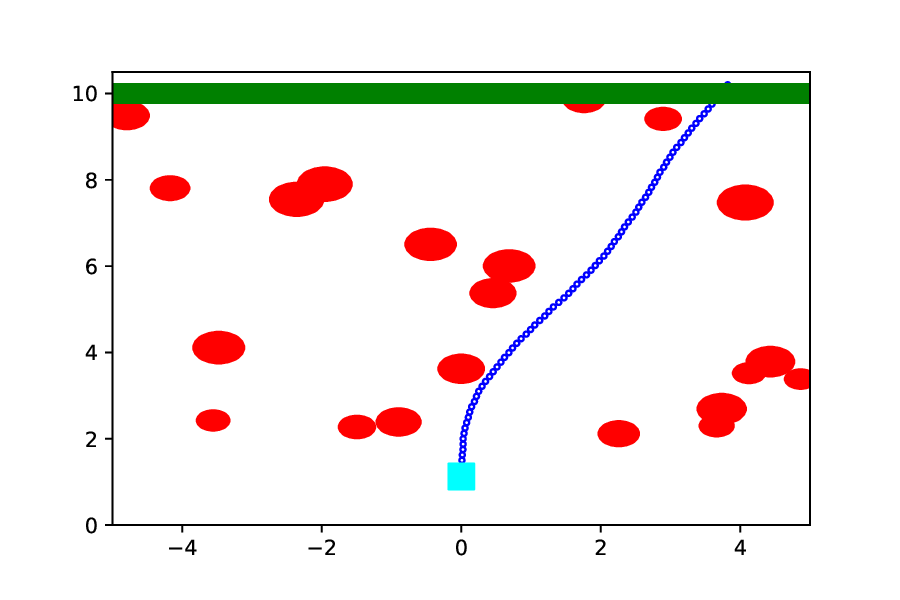}
 		\caption{Trajectory produced by $\theta^{[1]}_{k^{[1]}_{fs}}$}
 	\end{subfigure}
 	\hfil
 	\begin{subfigure}[t]{0.3\textwidth}
 		\centering
 		\includegraphics[width=1\textwidth]{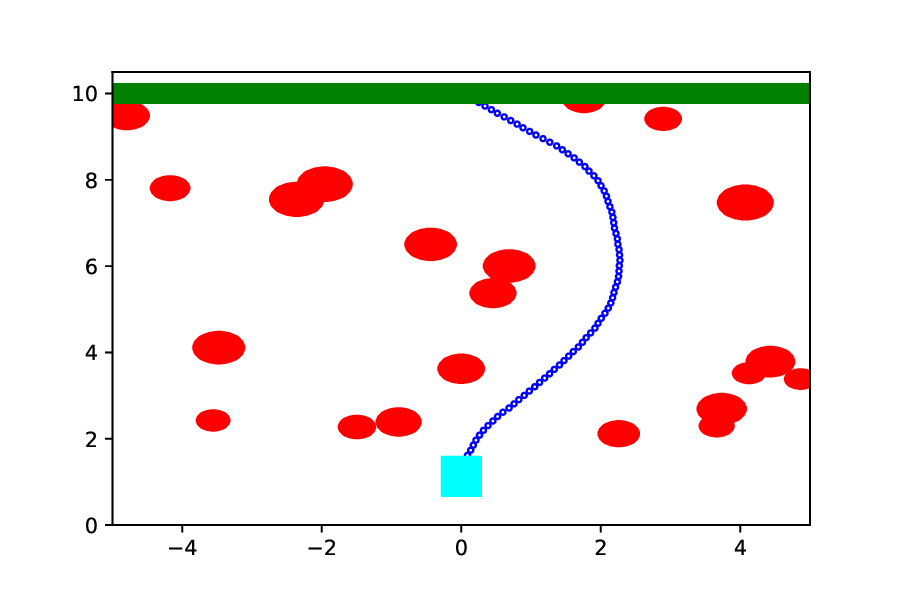}
 		\caption{Trajectory produced by $\theta^{[1]}_\infty$}
 	\end{subfigure}
 	\caption{Comparison between policies in  environment realization 5}
 	\label{fig: compare policy 5}
 \end{figure*}

\begin{figure*}
 	\centering
 	\begin{subfigure}[t]{0.3\textwidth}
 		\centering
 		\includegraphics[width=1\textwidth]{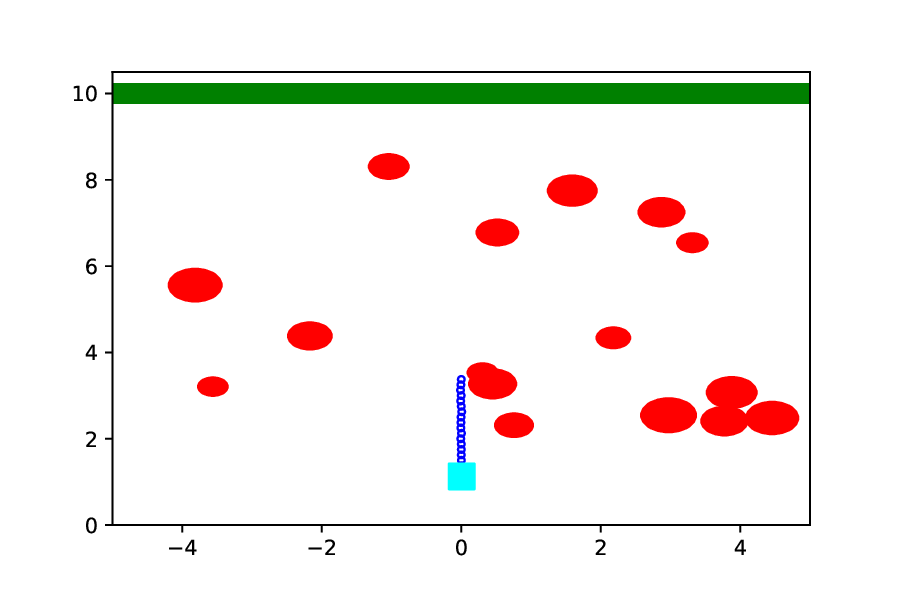} 
 		\caption{Trajectory produced by $\theta^{[1]}_0$}
 	\end{subfigure}
 	\hfil
 	\begin{subfigure}[t]{0.3\textwidth}
 		\centering
 		\includegraphics[width=1\textwidth]{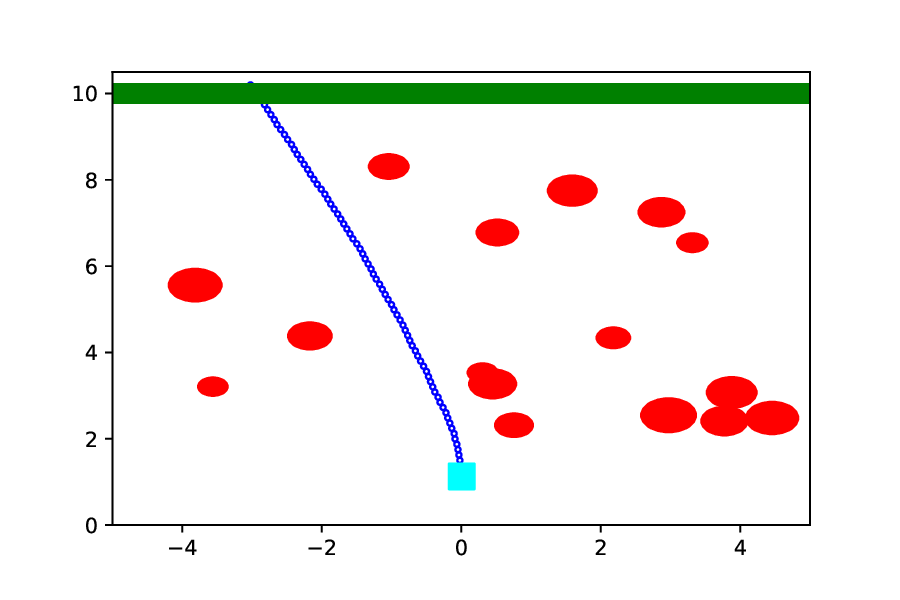}
 		\caption{Trajectory produced by $\theta^{[1]}_{k^{[1]}_{fs}}$}
 	\end{subfigure}
 	\hfil
 	\begin{subfigure}[t]{0.3\textwidth}
 		\centering
 		\includegraphics[width=1\textwidth]{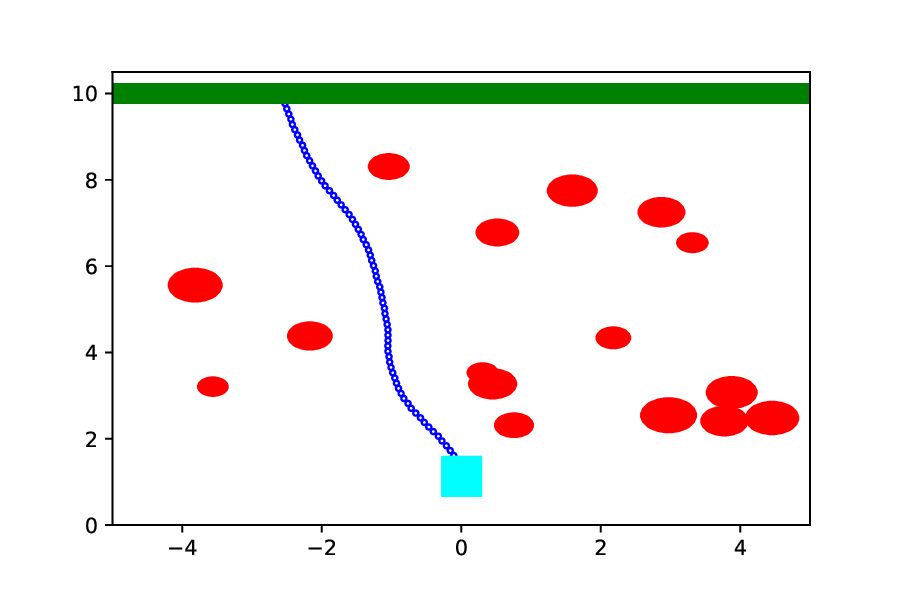}
 		\caption{Trajectory produced by $\theta^{[1]}_\infty$}
 	\end{subfigure}
 	\caption{Comparison between policies in  environment realization 6}
 	\label{fig: compare policy 6}
 \end{figure*}

Figures \ref{fig: compare policy 2} to \ref{fig: compare policy 6} show other realizations of the control policy in different environments.
 
\end{document}